\documentclass[%
 reprint,
 amsmath,amssymb,
 aps,
]{revtex4-1}

\usepackage{graphicx}% Include figure files
\usepackage{color}
\usepackage{dcolumn}% Align table columns on decimal point
\usepackage{bm}% bold math
\renewcommand\Re{\operatorname{Re}}
\renewcommand\Im{\operatorname{Im}}

\begin{document}

\preprint{APS/123-QED}

\title{Dyadic Green's function formalism for photo-induced forces in tip-sample nanojunctions}% Force line breaks with \\

\author{Faezeh Tork Ladani}
\affiliation{%
 Department of Electrical Engineering and Computer Science, University of California, Irvine
}%
\author{Eric Olaf Potma}%
 \email{epotma@uci.edu}
\affiliation{%
 Department of Chemistry, University of California, Irvine
}%\\
 \affiliation{Department of Electrical Engineering and Computer Science, University of California, Irvine}

\date{\today}% It is always \today, today,

\begin{abstract}
A comprehensive theoretical analysis of photo-induced forces in an illuminated nanojunction, formed between an atomic force microscopy tip and a sample, is presented. The formalism is valid within the dipolar approximation and includes multiple scattering effects between the tip, sample and a planar substrate through a dyadic Green's function approach. This physically intuitive description allows a detailed look at the quantitative contribution of multiple scattering effects to the measured photo-induced force, effects that are typically unaccounted for in simpler analytical models. Our findings show that the presence of the planar substrate and anisotropy of the tip have a substantial effect on the magnitude and the spectral response of the photo-induced force exerted on the tip. Unlike previous models, our calculations predict photo-induced forces that are within range of experimentally measured values in photo-induced force microscopy (PiFM) experiments. 

\begin{description}
\item[PACS numbers]
May be entered using the \verb+\pacs{#1}+ command.
\end{description}
\end{abstract}

\pacs{Valid PACS appear here}% PACS, the Physics and Astronomy
\maketitle

%\tableofcontents
%************************************************************************
%************************************************************************
%************************************************************************
%************************************************************************
\section{\label{sec:intro}Introduction}
Due to the conservation of linear momentum, optical fields can exert mechanical forces on the objects with which they interact. In the regime of (sub)-wavelength sized particles, these photo-induced forces are recognized as the time-averaged gradient and scattering force. Using laser light of a few mW, the photo-induced forces can manifest themselves in the pN range, which offers a means to control and manipulate objects at the micro- and nano-scales with light. This principle has been widely used in  optical tweezers and trapping studies of objects on the microscale~\cite{Ashkin1970,Ashkin1987}, optical binding of particles on the nanoscale~\cite{Dholakia2010}, and in plasmon-enhanced force manipulation of nano-particles~\cite{AriasGonzalez2003, Liu2010, Juan2011}.

Photo-induced forces also form the basis of photo-induced force microscopy (PiFM), where the local force exerted on an illuminated atomic tip is used as a contrast mechanism for imaging. This approach has been utilized to map out the local electric field of tightly focused laser beams~\cite{Huang2015}, propagating surface plasmon polaritons~\cite{JahngSPP2015} and localized surface plasmons~\cite{Kohoutek2011, Tumkur2016}. In addition, the optical binding force between the tip and a nearby polarizable object has been employed to generate photo-induced force maps of nanoparticles and molecules with nanometer scale resolution~\cite{Rajapaksa2010,Rajapaksa2011,JahngUltra2015,JahngACR2015,Nowake2016}. PiFM thus represents a promising tool for interrogating nanostructured samples with spectroscopic contrast at a spatial resolution that rivals other scan probe techniques. However, in order to extract meaningful and quantitative information from PiFM images, a better understanding is required of how photo-induced forces are translated into detectable mechanical motions of the atomic force microscopy (AFM) tip. One aspect of describing PiFM signals focuses on cantilever dynamics and frequency demodulation~\cite{Jahng2014,Jahng2016PRB}, whereas understanding the physics of the interactions in the tip-sample junction constitutes another open question. Developing a comprehensive and quantitative description of the forces relevant to PiFM has remained somewhat of a challenge.

Mechanical forces induced by the electromagnetic field are fully described by the surface integral of the time-averaged Maxwell's stress tensor (MST)~\cite{Novotny2012}. Because modeling of the full three-dimensional problem near the tip-sample junction is costly, simplifications are required to retrieve the essential physics at play. In a recent study, the photo-induced force was evaluated through the MST for an idealized spherical tip apex in a full wave simulation~\cite{Yang2016Raschke}.  This approach intrinsically includes multiple scattering mechanism between the tip and its image dipole or between the tip and an isolated particle. In the geometry examined, which did not include substrate effects, photo-induced forces were found that are substantially weaker than what is claimed in experimental work. Although full wave simulations are powerful, they do not lend themselves well for segmenting the problem into fundamental physical mechanisms that contribute to the overall observable force. For instance, it is unclear how much multiple scattering via the substrate contributes to the force, how gradient and scattering forces are independently affected by the tip-sample nanojunction, or how field gradients shape the magnitude of the force. For a deeper insight into the nanojunction photophysics, analytical models are indispensable and can guide the design of photo-induced force microscopes based on the relevant mechanisms that constitute the force. 

Simple analytical models based on the dipolar approximation of the tip's polarizability offer clear insights into the behavior of the photo-induced force exerted on the tip~\cite{Chaumet2000,NietoVesperinas2004,Dholakia2010,Rajapaksa2010,Jahng2016spie}. In particular, dipole-dipole based tip-sample interactions predict a distance dependent gradient and scattering force that closely resembles experimental observations~\cite{Jahng2014,Jahng2016PRB}. This model provides a useful physical picture for the tip-sample interaction, yet it neglects the multiple scattering mechanisms between the tip and sample through the substrate. In addition, the predicted forces are generally smaller than what is observed experimentally in PiFM. It is known that multiple scattering effects are important for practical tip-sample geometries, as was recently emphasized in a theoretical study of the local fields in the nanojunction as it pertains to tip-enhanced Raman scattering~\cite{Zhang2015}. A recent theoretical analysis of photo-induced forces between spherical nanoparticles, which includes sample-substrate interactions, provides a comprehensive formalism for quantitatively assessing forces at the nanoscale~\cite{Salary2016}. This latter model can be applied to describe the forces measured in PiFM, although the formalism is not designed to naturally discriminate between separate scattering mechanisms or to include the anisotropic geometry of the AFM tip. 

In this work, we present an alternative description of the photo-induced force in the dipolar approximation by adopting a dyadic Green's function (GF) approach for expressing the local fields near a planar multilayer substrate. Unlike previous models, this description allows analytical expressions that enable a closer look at the multiple scattering pathways that shape the field and its gradient, which jointly define the photo-induced force. Through the application of this Green's function formalism, we show that the magnitude of the force in typical PiFM measurements is largely dictated by the steep field gradients in the tip-sample junction, which are a direct consequence of multiple scattering between the tip, sample and substrate. We also establish that scattering pathways via the substrate make a significant contribution to the photo-induced force in PiFM measurements, and cannot be ignored. In addition, we show that by including the anisotropy of the tip's polarizability, photo-induced forces are predicted that compare well with previously reported experiments.   

%************************************************************************
%************************************************************************
%*****************        Theory          *******************************
%************************************************************************
%************************************************************************

\section{\label{sec:theory}Optically induced forces using the dyadic Green's function }

\subsection{\label{sec:basics} Photo-induced force and electrodynamics of the illuminated nanojunction using Green's function formalism}
In this Section we will discuss the photo-induced force exerted on a tip with the help of a dyadic Green's function formalism for a multilayer planar substrate. We assume that the tip's apex can be modeled effectively in the dipolar limit, i.e. higher order multipoles are assumed to be negligible. The tip's response to a time-harmonic field, using the \(e^{-i\omega t}\) convention, is described by the complex electronic polarizability $\bar{\boldsymbol\alpha}$, which gives rise to a dipole moment \(\mathbf{p}\) in the tip, where \(\mathbf{p}=\bar{\boldsymbol\alpha} \cdot \mathbf{E}\). We will indicate tensor quantities in bold with an overhead bar. The net time-averaged force experienced by the dipole in a spatially inhomogeneous field can be expressed as:~\cite{Novotny2012}
\begin{equation}
\left<\mathbf{F}(t)\right>=\frac{1}{2}\displaystyle \sum_{l}\Re\left\{p^{l*} \nabla E^l \right\}
\label{eq:force1}
\end{equation}
where $l=x,y,z$ labels the polarization components along each of the cartesian coordinates. In case the  polarizability is isotropic, as defined by its diagonal elements $(\alpha_{xx}=\alpha_{yy}=\alpha_{zz})$~\cite{Jahng2014,Yang2016Raschke}, the force in Eq. (\ref{eq:force1}) can be expanded as:
\begin{equation}
\left<\mathbf{F}(t)\right>=\frac{\alpha^\prime}{2} \displaystyle \sum_{l}\Re\left\{E^{l*} \nabla E^l \right\}+\frac{\alpha^{\prime\prime}}{2} \displaystyle \sum_{l}\Im\left\{E^{l*} \nabla E^l \right\}
\label{eq:force2}
\end{equation}
where $\alpha=\alpha^{\prime} + i\alpha^{\prime\prime}$, and \(E^l\) and \(\nabla E^l\) are the local electric field and the gradient of the field for the different polarization components $l$ of the field. In Eq.~(\ref{eq:force2}), the first term is called the gradient force since $\frac{\alpha^\prime}{2}\sum_{l}\Re\left\{E^{l*} \nabla E^l \right\}=\frac{\alpha^\prime}{2} \nabla|\mathbf E|^2$ has zero circulation. The second term is often referred to as the scattering force.
Note that separating the photo-induced force into the gradient and scattering forces is possible when the particle of interest is isotropic with a scalar polarizability. For an anisotropic particle such as the tip, Eq.~(\ref{eq:force2}) is not generally valid, and we will use Eq.~(\ref{eq:force1}) to calculate the force instead. 

\begin{figure}[ht]
	\begin{center}
		\includegraphics[width=0.92\columnwidth]{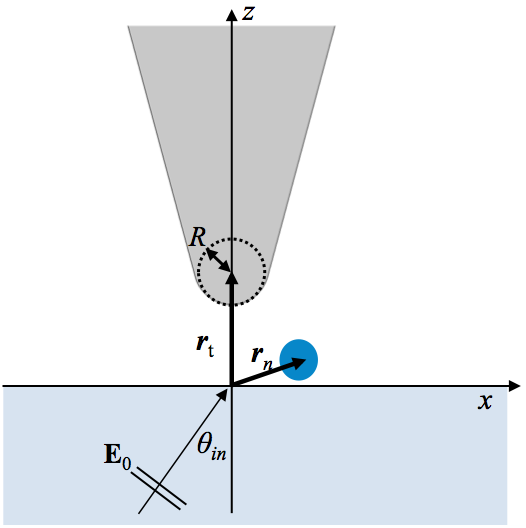}
	\end{center}
	\caption{PiFM setup for probing photo-induced forces. An AFM tip and a nanoparticle (NP) are located above the substrate, and illuminated from below through the substrate. }
	\label{fig:Figure1}
\end{figure}

In the following, we will denote the time averaged force, \(\left<\mathbf{F}(t)\right>\) as \(\mathbf{F}\) without a time dependence and we note that all other parameters discussed in the paper are expressed in the frequency domain. The relevant geometry is shown in Figure~\ref{fig:Figure1}, where the tip and any nearby polarizable nanoparticle (NP) are located above a (multilayer) planar substrate and illuminated with a plane wave of amplitude $E_0$ that is incident at the angle $\theta_{in}$ relative to the surface normal. The electric field above the substrate surface \(z>0\) in the absence of the tip is the transmitted plane wave with a wave vector and amplitude defined by the Fresnel equations.

We will call the transmitted field the background field \(\mathbf{E}_b\). The subscripts \(t\) and \(n\) denote the location of the tip and NP, respectively, and will be used to indicate the  region of interest for \(\mathbf{E}_b\). By applying the dipolar approximation for the tip and other scattering nanoparticles in the system, we establish the electrodynamic model for \(\mathbf{p}_t\) and \(\mathbf{p}_n\), the induced dipole moment in the tip and the NP, respectively. For the general configuration of the two scatterers in air for \(z>0\), as shown in Fig.~\ref{fig:Figure1}, the electric field at \(\mathbf{r}_t\) is the sum of (i) the background field at the location of the tip (\(\mathbf{E}_{b;t}\)), (ii) the self interaction of \(\mathbf{p}_t\) through the substrate and (iii) the field scattered by \(\mathbf{p}_n\). Using dyadic Green's functions (GF), we can write the following: 
\begin{equation}
\mathbf{E}_t=\mathbf{E}_{b;t}+\bar{\mathbf{G}}_{sc;t,t}\cdot\mathbf{p}_t+\bar{\mathbf{G}}_{t,n}\cdot\mathbf{p}_n 
\label{eq:EtotDyadic}
\end{equation}
where \(\bar{\mathbf{G}}_{sc;t,t}=\bar{\mathbf{G}}_{sc}(\mathbf{r}_t, \mathbf{r}_t)\) is the scattering GF which accounts for the self interaction of the tip via the substrate, and the total GF, \(\bar{\mathbf{G}}_{t,n}=\bar{\mathbf{G}}_{sc}(\mathbf{r}_t, \mathbf{r}_n)+\bar{\mathbf{G}}_{0}(\mathbf{r}_t, \mathbf{r}_n)\), is the sum of the free space and the scattering 
GFs, which account for the direct and indirect (through the substrate) coupling of the tip and the NP. Assuming that both scatterers are above the planar multilayered medium \(z>0\), the scattering GF for an observer and source located at \(\mathbf{r}\) and \(\mathbf{r}^\prime\), respectively, can be written as the integral of scattered plane waves in the angular spectrum representation~\cite{Kong1990}:
\begin{align} \nonumber
\bar{\mathbf{G}}(\mathbf{r}, \mathbf{r^\prime}) = \frac{i}{8\pi^2} \iint dk_x dk_y \frac{1}{k_{0z}}\times \qquad\\
\left[R^{TE} \hat{\mathbf{e}}(k_{0z}) \hat{\mathbf{e}}(-k_{0z}) + R^{TM} \hat{\mathbf{h}}(k_{0z}) \hat{\mathbf{h}}(-k_{0z}) \right] e^{i \mathbf{k} \cdot \mathbf{r}} e^{-i \mathbf{K} \cdot \mathbf{r^\prime}}
\label{eq:GscIntDef}
\end{align}
Here \noindent\(\hat{\mathbf{e}}\), \(\hat{\mathbf{h}}\), \(\hat{\mathbf{k}}\) are the unit vectors for the electric field, the magnetic field and the wavevector, which form an orthonormal system for representing any TE/TM polarized wave. The wavevector \(\mathbf{k}\) describes propagation along \(+z\) and is written as $\mathbf{k}=\hat{\mathbf{x}}k_x+\hat{\mathbf{y}}k_y+\hat{\mathbf{z}}k_{0z}$.  Similarly, $\mathbf{K}$ is the wavevector for propagation along \(-z\) and is defined as \(\mathbf{K}=\hat{\mathbf{x}}k_x+\hat{\mathbf{y}}k_y-\hat{\mathbf{z}}k_{0z}\), where the lateral components of the wave vector are indicated as \(k_x\) and \(k_y\), and the longitudinal component is written as $k_{0z}=\sqrt{k^2_0-k^2_x-k^2_y}$ for \(z>0\). The intrinsic wavenumber $k_0$ is defined as $\sqrt{\varepsilon_r\mu_r}(\omega/c)$. \(R^{TE}\) and \(R^{TM}\) are the transverse electric and transverse magnetic Fresnel coefficients, respectively. For more details on the scattering GF see \cite{Kong1990}. 

Expanding the dipole moments \(\mathbf{p}_j=\bar{\boldsymbol\alpha}_j \cdot \mathbf{E}_j\) with $j=t,n$ in Eq.~(\ref{eq:EtotDyadic}), we arrive at the following expression for the local electric field at the tip:
\begin{equation}
\mathbf{E}_t=\left[\bar{\mathbf{S}}_{tt}-\bar{\mathbf{M}}_{tn} \bar{\mathbf{S}}^{-1}_{nn} \bar{\mathbf{M}}_{nt} \right]^{-1} (\mathbf{E}_{b;t}+ \bar{\mathbf{M}}_{tn} \bar{\mathbf{S}}^{-1}_{nn} \cdot \mathbf{E}_{b;n})
\label{eq:EtotDyadic2}
\end{equation}
\noindent where \(\bar{\mathbf{S}}_{jj}=\bar{\mathbf{I}}-\bar{\mathbf{G}}_{sc;jj}\cdot \bar{\boldsymbol\alpha}_j\) is the self interaction matrix for scatterer \textit{j} and \(\bar{\mathbf{M}}_{jk}=\bar{\mathbf{G}}_{jk}\cdot \bar{\boldsymbol\alpha}_k\) is the mutual interaction matrix. In the expression of $\bar{\mathbf{M}}_{jk}$, the total Green's function \(\bar{\mathbf{G}}_{jk}\) tracks the influence of scatterer \textit{k} on the fields at the location of scatterer \textit{j}. Note that the interaction matrices are unitless. Although Fig~\ref{fig:Figure1} shows a one layer substrate, this formulation is general for a multi-layered substrate with different permittivities in the (\(z<0\)) region. In all cases, appropriate reflection and/or transmission coefficients should be used for calculating the scattering GF and \(\mathbf{E}_b\).

\subsection{\label{sec:parameters}Computational details}
For the isotropic model of the tip, we will assume that the tip behaves as a polarizable spherical particle with an effective dipolar response~\cite{Novotny1998, Novotny2012,Ichimura2007, Jahng2014}. In this case, the tip apex characterized by a $R=30$ nm radius, similar to the experimental value reported in ~\cite{JahngSPP2015}. The tip is composed of gold and the template-stripped gold permittivity described in \cite{Olmon2012} is used in our calculations. The scalar polarizability of the tip is calculated with the aid of the Mie coefficients~\cite{Bohren1998, Campione2011} for a nanosphere of the apex dimension. For the anistropic tip, we will use the formalism described in Appendix \ref{App:anisotropic}. The substrate is glass with a refractive index of 1.5. We will consider the forces in the vicinity of the bare glass surface as well as a glass surface covered with a 45 nm gold film, using the refractive index reported in \cite{McPeak2015}. The incident field is a \textit{p} polarized plane wave with \(E_0=10^6~ V/m\) illuminated at \(\theta_{in}\). In both cases, for the bare glass surface as well as the glass surface covered with gold, \(\theta_{in}\) is set to $=43.65^o$ to produce an evanescent background field for \(z>0\). Note that this angle is at the Kretschmann angle for the gold covered substrate, thus launching a surface plasmon polariton at the gold/air interface. Unless otherwise noted, we will focus on $F_z$, which is the photo-induced force directed along the $z$-axis. 

To compute the force, the GF in Eq. (\ref{eq:GscIntDef}) and its spatial derivatives are evaluated numerically. Once these functions are determined, the field and field gradient can be obtained analytically, which subsequently allow the calculation of the photo-induced force through Eq.~(\ref{eq:force1}). In the following Sections, we will provide explicit expressions for the fields and field gradients along $z$ that are relevant to the different sample geometries considered here.  

%************************************************************************
%************************************************************************
\subsection{\label{sec:tipSubs}Field versus field gradient}
Before studying the behavior of the photo-induced force in detail, we will first examine two key aspects of the force, namely the magnitude of the field and the field gradient. For this discussion, we will consider the scenario of an isotropic gold tip in the vicinity of an glass/air or gold/air planar interface. This corresponds to the situation sketched in Figure \ref{fig:Figure1} without the presence of the NP.  We may thus set \(\bar{\boldsymbol\alpha}_n=\bar{\mathbf{0}}\) in Eq.~(\ref{eq:EtotDyadic2}). In this case, the expression for the electric field at the tip simplifies considerably. The components of the local electric field at the tip can then be written as:
\begin{equation}
E^l_t=\frac{E^l_{b;t}}{S^{ll}_{tt}} ~~ l=x,y,z
\label{eq:Etotscalar1}
\end{equation}
 \noindent where \(S^{ll}_{tt}=1-\alpha^{ll}_{t}G^{ll}_{sc;tt}\), \(E^l_{b;t}=E^l_{b0}e^{ik_zz_t}\) is the background field at the location of the tip, and \(k_z\) is the wavenumber along \textit{z} for \(z>0\). Because we will consider an evanescent background field, we can use \(k_z=i\gamma\) to represent the field amplitude decay along \textit{z} for \(z>0\). We assume that the polarizability of the tip is not dependent on the tip-surface distance, therefore using Eq.~(\ref{eq:Etotscalar1}), the field gradient along \textit{z} direction is 
\begin{equation}
\frac{\partial}{\partial z}E^l_t=E^l_t \left[-\gamma+ \frac{\alpha^{ll}_{t}}{S^{ll}_{tt}} \frac{\partial G^{ll}_{sc;tt}}{\partial z} \right].
\label{eq:gradE2}
\end{equation}
The photo-induced force in the longitudinal direction can then be written as:
%\begin{equation}
%F_z= \frac{1}{2} \displaystyle \sum_{l} \left|E^l_t\right|^2 \left(-\gamma\alpha_t^{ll\prime }+\left|\frac{\alpha^{ll}_{t}}{S^{ll}_{tt}}\right|^2 \Re\left\{ \frac{\partial G^{ll}_{sc;tt}}{\partial z}S^{*ll}_{tt} \right\}\right) 
%\label{eq:FzScal}
%\end{equation}
\begin{equation}
F_z= \frac{1}{2} \displaystyle \sum_{l} \left|E^l_t\right|^2 \left(-\gamma\alpha_t^{ll\prime }+\left|\alpha^{ll}_{t}\right|^2 \Re\left\{\frac{1}{S^{ll}_{tt}} \frac{\partial G^{ll}_{sc;tt}}{\partial z} \right\}\right) 
\label{eq:FzScal}
\end{equation}
The magnitude of the photo-induced force depends on two terms: one term that scales solely with the magnitude of the field components, and a second term that also shows a dependence on the field gradient. We will examine which quantity, the field or the field gradient, is the dominant factor in the typical PiFM scenario considered here. Both the field and its gradient depend on the self interaction term \(S^{ll}_{tt}=1-\alpha^{ll}_{t}G^{ll}_{sc;tt}\), which appears in the denominator of in Eqs.~(\ref{eq:Etotscalar1}) and~(\ref{eq:gradE2}). This self interaction term contains a resonance condition when \(\Re\{\alpha^{ll}_{t}G^{ll}_{sc;tt}\}\approx 1\) and \(\Im\{\alpha^{ll}_{t}G^{ll}_{sc;tt}\}\approx 0\), which can be associated with a spatial resonance that gives rise to the local enhancement of the field. A detailed description of this phenomenon for a dipolar scatterer beside a plasmonic nanosphere is presented in \cite{FTL2014}.

To study how the field and the field gradient are affected by the process of multiple scattering, we define two quantities, namely the field enhancement factor \(FE^{ll}\), which is normalized to the background field, and the normalized gradient specific enhancement factor \(GSE^{ll}\), as 
\begin{align} \nonumber
FE^{ll}=\frac{1}{|S^{ll}_{tt}|}\qquad\\
GSE^{ll}=\frac{\left|\alpha^{ll}_{t}\right|^2}{-\gamma\alpha^{ll\prime}_t} \Re \left\{ \frac{1}{S^{ll}_{tt}}\frac{\partial G^{ll}_{sc;tt}}{\partial z_t} \right\}
\label{eq:EnhFactors}
\end{align}

The field scales with $FE^{ll}$, as determined by the spatial resonance described by  $S^{ll}_{tt}$. The field gradient, on the other hand, has two contributions, one that scales as $FE^{ll}$ and the second that scales as \(GSE^{ll}\). The latter term is the gradient specific term, which depends more strongly on the process of multiple scattering than the $FE^{ll}$ term. Hence, it can be expected that field gradient effects grow in importance when multiple scattering effects become more significant. 

\begin{figure}[ht]
	\begin{center}
		\includegraphics[width=0.92\columnwidth]{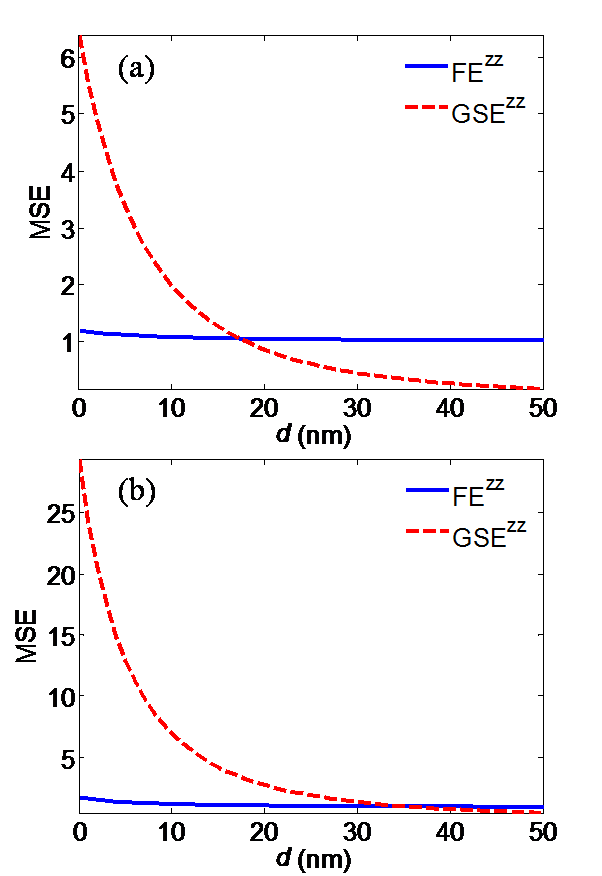}
	\end{center}
	\caption{Enhancement factor due to multiple scattering effect (MSE), comparing the distance dependence of the normalized field enhancement ($FE^{zz}$) factor and the normalized gradient specific enhancement ($GSE^{zz}$) factor using illumination at 450 THz. (a) Tip in the vicinity of a glass substrate. (b) Tip in the vicinity of a glass substrate covered with a 45 nm gold film.} 
	\label{fig:FE_GSE}
\end{figure}

The distance dependence of $FE^{zz}$ and $GSE^{zz}$ are displayed in Figure \ref{fig:FE_GSE}. In panel \ref{fig:FE_GSE}(a), the enhancement factors are shown for the force measured in the vicinity of the bare glass surface. With reference to Figure \ref{fig:Figure1}, the surface to surface distance $d$ is defined as $z_t-R$. We see that $FE^{zz}$ shows only limited enhancement as a function of distance. A small enhancement is seen for shorter distances ($d<10$ nm), due to the spatial confinement of the field in the tip-sample junction as described by $S^{zz}_{tt}$. The gradient specific enhancement factor, on the other hand, shows a much more prominent distance dependence. Since both quantities are normalized, we may compare the values of $FE^{zz}$ and $GSE^{zz}$ as the tip-sample distance is shortened. It is clear that the enhancement in the gradient specific contribution is significantly higher than the enhancement in the field, as expected for shorter distances when the effects of multiple scattering are becoming more relevant. Note that these results are relevant to the dipole approximation considered here, which may underestimate the extend of field confinement effects. Even stronger field and field gradient effects are expected when multipoles are included~\cite{FTL2014}. 

In Figure \ref{fig:FE_GSE}(b) a similar comparison is shown between $FE^{zz}$ and $GSE^{zz}$ for the tip in the vicinity of the thin gold film. Due to stronger permittivity contrast at the gold/air interface compared to glass/air interface, the scattering GF for a gold/glass substrate has higher amplitude. Consequently, the field enhancement between the gold surface and the tip's apex is more prominent. The field gradient specific enhancement is even more significant, especially for distances $d<10$ nm. In this scenario, the behavior of the photo-induced force is largely dictated by the gradient field effects. The latter observation is a direct consequence of the enhanced multiple scattering.

%************************************************************************
%************************************************************************
%********************* Numerical evaluation *****************************
%************************************************************************

\section{\label{sec:tiponlyCal}Behavior of the photo-induced force}
In this Section, we present calculations of the photo-induced force for several typical experimental geometries that are relevant to PiFM imaging~\cite{Jahng2016spie}. First, we will consider the photo-induced force exerted on the tip over a bare glass substrate. Second, we will highlight the photo-induced force when the tip is driven by a surface plasmon polariton on a thin gold film. Third, we will discuss the magnitude and behavior of the photo-induced force when the anisotropic polarizability of the tip is taken into account. Lastly, we will focus on the photo-induced forces between the tip and a polarizable nanoparticle on the substrate. Our goal is to show that the process of multiple scattering and tip anisotropy are important and can modify the magnitude and spectral dependence of the force considerably.

%************************************************************************
%************************************************************************
\subsection{\label{sec:Isopol}Photo-induced force over a glass substrate}
\begin{figure}[ht]
	\begin{center}
		\includegraphics[width=0.92\columnwidth]{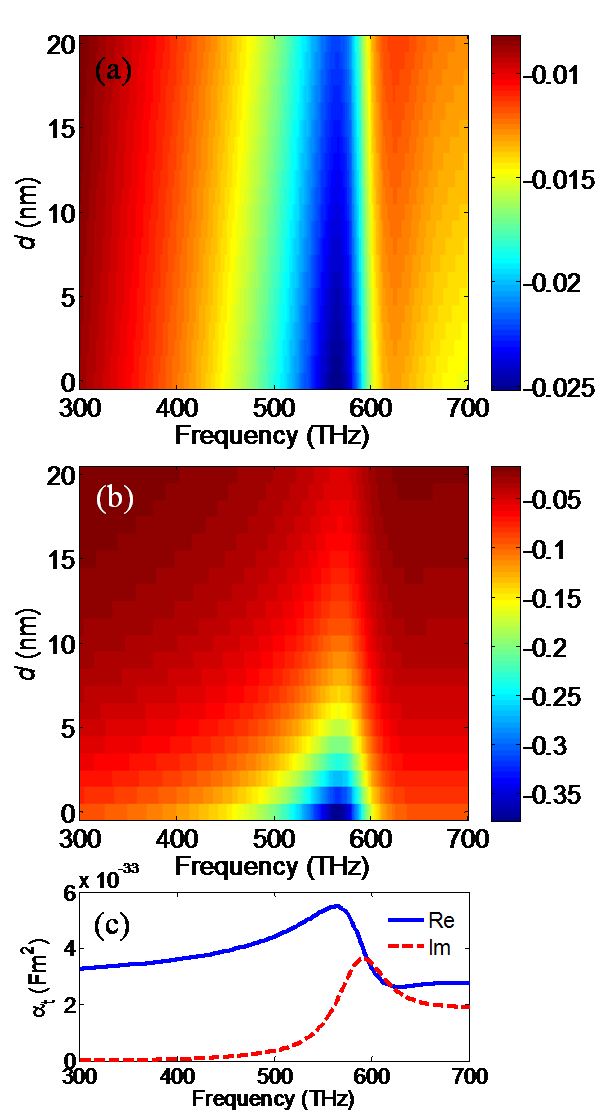}
	\end{center}
	\caption{Time-averaged photo-induced force $F_z$ exerted on the gold tip as a function of frequency and tip-substrate distance $d$. (a) $F_z$ without including multiple-scattering. (b) $F_z$ calculated with multiple-scattering included. Force is given in picoNewton (pN).(c) Frequency dependence of the real (blue) and imaginary (red) parts of the tip's polarizability.}
\label{fig:Figure3}
\end{figure}

We first examine the force when an isotropic tip is positioned over a bare glass substrate under evanescent illumination conditions. Figure~\ref{fig:Figure3} shows the frequency and distance dependence of the computed \(F_z\) exerted by the evanescent background field on the tip in pN. Note that the minimum force sensitivity depends on the specifics of the AFM system used for force mapping and can vary from tens of fN to hundreds of fN under ambient conditions~\cite{Kohoutek2011, Wickramasinghe2011, Yang2016Raschke}. For our system this value is about 0.1 pN \cite{JahngSPP2015}. In Fig.~\ref{fig:Figure3}(a) the magnitude of the force is shown without the effect of multiple scattering, calculated as  \(F_z=\frac{\alpha^\prime_t \gamma}{2}|\mathbf{E}_{b;t}|^2\) where \(\alpha^\prime_t\) is the dispersive part of the tip polarizability and \(\gamma\) is the evanescent wavenumber of the background field (see Appendix A for details). In Fig.~\ref{fig:Figure3}(b) the effect of multiple scattering effect is included. The spectral dependence of the dispersive and dissipative parts of the tip's polarizability are indicated by the solid and dashed lines, respectively, in panel~\ref{fig:Figure3}(c).

\begin{figure}[ht]
	\begin{center}
		\includegraphics[width=0.92\columnwidth]{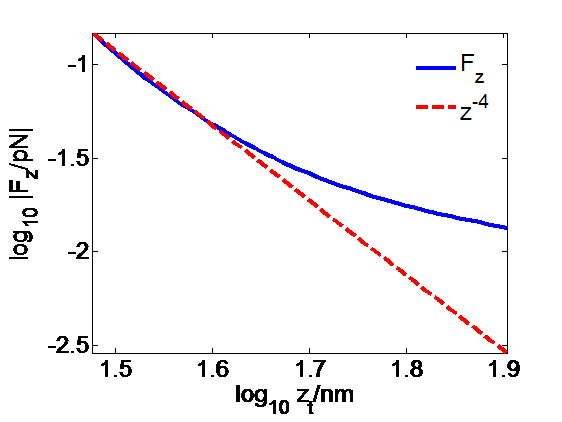}
	\end{center}
	\caption{Time-averaged photo-induced force $F_z$ exerted on the  tip as a function of tip position $z_t$. Results from the scattering Green's function analysis (blue solid line) are shown in addition to a $z^{-4}$--dependence (red dashed line). The agreement between the two models is good up to $\sim9$ nm.}
\label{fig:z_dep_comp}
\end{figure}

In both cases, the photo-induced force \(F_z\) is a negative (attractive) force, meaning that the tip is pulled toward the surface. In addition, the frequency dependence of the force closely traces the  dispersive part of the polarizability, reaching its maximum value near the plasmonic resonance of the tip apex. However, there are also clear differences between Fig.~\ref{fig:Figure3}(a) and (b). Without multiple-scattering, the distance dependence of the photo-induced force is shallow, largely dictated by the $z$-dependence of the evanescent background field. On the other hand, with multiple-scattering included, the force not only increases in magnitude by about 15-fold, but also becomes more localized. We see that the force is confined to shorter tip-substrate distances of up to 10 nm (Fig.~\ref{fig:Figure3}(b)), displaying a much sharper distance dependence. This feature is a direct consequence of the coupling between the tip and the substrate, which grows nonlinearly as the tip approaches the substrate surface. The observed distance dependence of the force mimics the interaction between the tip dipole and its image in the substrate. The latter scenario in PiFM is usually referred to as the image dipole force, which, using a simple analytical model based on dipole-dipole interaction, results in a $\sim z^{-4}$ dependence for the gradient force when only the $E_z$ component of the field is considered~\cite{Jahng2016spie}. Our formalism includes all field components and generalizes the force for different illumination conditions. Figure \ref{fig:z_dep_comp} shows the distance dependence of the force for illumination at 450 THz using our GF model. We find that the photo-induced force is still highly confined, but does not follow exactly a $z^{-4}$--dependence over the entire range of frequencies and distances examined here. In Figure \ref{fig:z_dep_comp}, for shorter distances up to $\sim9$ nm, the GF-based calculation is in close agreement with a $z^{-4}$--dependence, while the results deviate for larger tip-sample distances. The difference largely reflects the effect of the different illumination conditions. Whereas the $z^{-4}$ dependence is derived for propagating light, here we consider the force in an evanescent illumination field. 

%************************************************************************
%************************************************************************
\subsection{\label{sec:goldsurface}Photo-induced force over a gold surface with surface plasmon polarition excitation}

\begin{figure}[ht]
	\begin{center}
		\includegraphics[width=0.92\columnwidth]{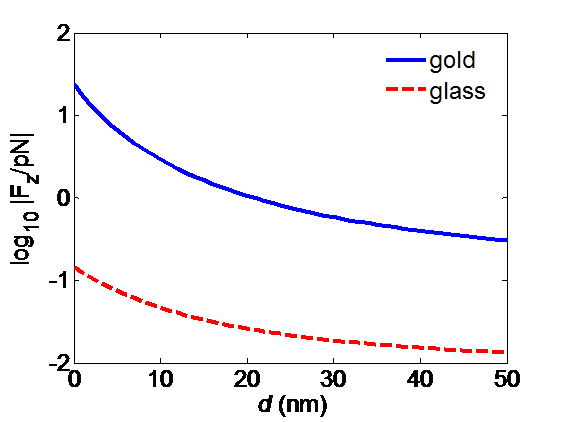}
	\end{center}
	\caption{Distance dependence of the photo-induced force exerted on the tip when exciting an SPP mode (\(\theta_i=43.65^o\)) at the gold/air interface (blue line). The red dashed line corresponds to the same measurement on the bare glass substrate. Illumination is at 450 THz.}
\label{fig:SPPvsGlass}
\end{figure}

We next consider the photo-induced force for the case of a gold tip positioned over a glass substrate covered with a 45 nm gold thin film. As was clear from the discussion in Section \ref{sec:tipSubs}, compared to the bare glass substrate, the field and its gradient under SPP excitation are more significant in the nano-junction formed by the tip and the gold surface. This translates into a higher magnitude of the force. In   Fig.~\ref{fig:SPPvsGlass}, the tip-sample distance dependence of $F_z$ is compared for the case of the bare glass surface and the thin gold film. As expected, the force in the case of the gold surface is stronger than what is measured over the glass surface. Under the conditions examined here, we find a force that is approaching 24 pN as the tip gets closer to the gold surface, more than two orders of magnitude stronger than the force experienced by the tip over the glass substrate. The magnitude of the computed force roughly complies with PiFM measurements of SPP fields on a gold surface~\cite{JahngSPP2015}.

%************************************************************************
%************************************************************************
\subsection{\label{sec:AnisoPol}Photo-induced force at a tip with anisotropic polarizability}
\begin{figure}[ht]
	\begin{center}
		\includegraphics[width=0.92\columnwidth]{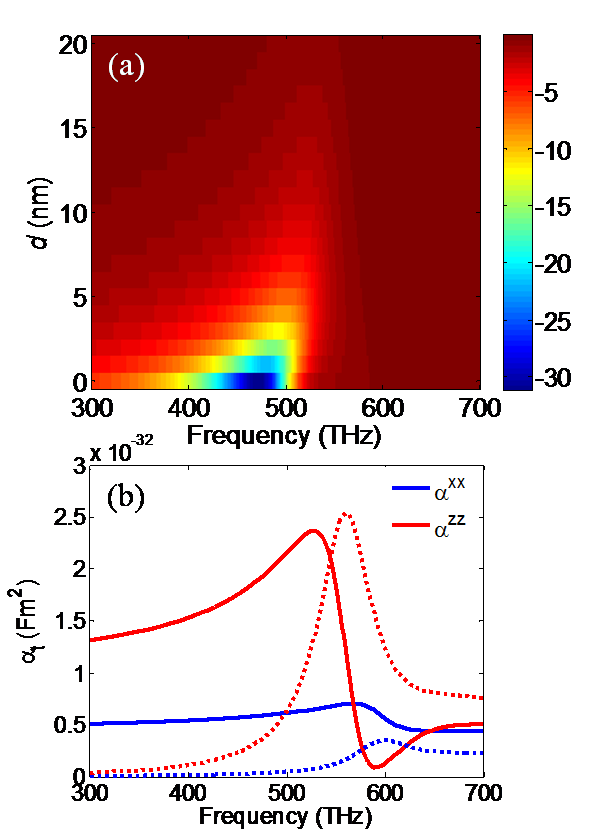}
	\end{center}
	\caption{Effect of anisotropic polarizability of the tip. (a) \(F_z\) as a function of distance and frequency, using $AF=2$. The tip is place over a glass substrate, using \(\theta_i=43.65^o\). (b) Spectral dependence of the real (solid line) and imaginary (dotted line) parts of the tip's polarizability tensors $\alpha_{zz}$ (red) and $\alpha_{xx}$ (blue) for $AF=2$.}
	\label{fig:AF_2}
\end{figure}

The spherical polarizability model is widely used to model an AFM tip. Yet, it is well known that an actual tip exhibits an anisotropic response, which gives rise to a stronger field enhancement in the longitudinal polarization direction of the excitation field \cite{Novotny1997,Bouhelier2003,Novotny2006,Novotny2012}. 
The stronger optical response along the axial dimension gives rise to photo-induced forces that are stronger than what is predicted in the spherical polarizability model. In this section, we investigate the possible effects of tip anisotropy on $F_z$ when adopting an anisotropic description of the tip's polarizability. We use a polarizability model for a virtual prolate particle~\cite{Link1999,Imura2011}, taking into account losses due to radiation~\cite{Anger2006}, see Appendix B for details. The anisotropy factor, \(AF\), is defined as the aspect ratio of the major to minor axis of the virtual prolate, as a perturbation to the spherical apex. It should be mentioned that the location of the effective induced dipole moment of the tip is considered the same as the isotropic model, for the sake of comparison. In Figure~\ref{fig:AF_2}(a) we show the calculated frequency and distance dependence of the photo-induced force \(F_z\) for \(AF=2\) when the tip is placed over the bare glass substrate. The spectral dependence of the longitudinal and transverse polarizabilities is plotted in Figure~\ref{fig:AF_2}(b), clearly showing the enhanced response of the tip along the axial direction. The enhancement of the polarizability by about 5 times in the longitudinal direction that we consider here is within empirical ranges for a gold tip~\cite{Novotny2006}.

\begin{figure}[ht]
	\begin{center}
		\includegraphics[width=0.92\columnwidth]{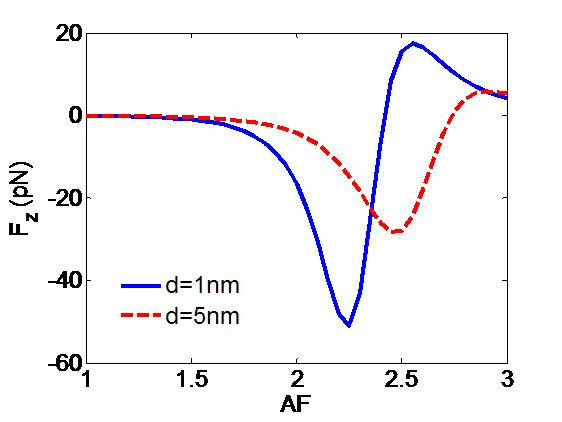}
	\end{center}
	\caption{Photo-induced force as a function of the tip anisotropy factor $AF$, shown for two different tip-sample distances. Substrate is a bare glass surface and illumination frequency is 450 THz.}
	\label{fig:FzAF}
\end{figure}

Compared to the isotropic tip, discussed in Fig.~\ref{fig:Figure3}(a),  Fig.~\ref{fig:AF_2}(a) underlines that the photo-induced force can grow to substantial values when the tip's anisotropy is included. We find that attractive forces up to 30 pN can be reached for $AF=2$, an increase of almost two orders of magnitude compared to the calculation in Fig.~\ref{fig:Figure3}(a). The much stronger forces are a direct consequence of not only the enhanced local field but also of the steeper field gradient along $z$. 
In addition to an enhancement of $F_z$, we see that the maximum force is now found at lower frequencies. This frequency shift is not only related to the shifted spectral resonance for $\alpha_{zz}$, but also to the spatial resonances that manifest themselves in the tip-sample junction. To illustrate the latter point, we show the magnitude of $F_z$ at 450 THz as a function of the anisotropy factor in Figure~\ref{fig:FzAF}. As $AF$ is increased, the attractive photo-induced force grows to more than 50 pN before decreasing and turning into a repulsive force. This somewhat unexpected dependence can be explained through the concept of spatial resonance. The field enhancement is a function of the magnitude of the polarizability, producing a resonance when the scattered field amplitude matches the background field and interferes with it constructively. The (real) gradient of this spatial resonance in the field undergoes a sign change, resulting in a dispersive lineshape, and turning an attractive force into a respulsive force as the tip's responsiveness continues to grow. The observed sign reversal of the force is remarkable, and underlines that, unlike most optical effects in the tip-sample junction, the force does not only scale with the field amplitude but also (very sensitively) with its gradient.

%************************************************************************
%************************************************************************

\subsection{\label{sec:TipSample}Photo-induced force between a tip and a nanoparticle}
Analyzing the isolated tip above a multilayer surface is useful for near field probing of surface fields. In many PiFM applications, however, we are interested in imaging nanostructures on top of the substrate, such as sketched in Figure \ref{fig:Figure1}. We will first explore the behavior of the confined fields in this configuration in the context of the scattering Green's function. In particular, we aim to show the effects of indirect interactions through multiple scattering via the substrate compared to direct scattering via the nanoparticle. 

To gain insight in the local field, Eq.~(\ref{eq:EtotDyadic2}) can be reduced to a set of scalar equations for the electric field at the tip in the limit of a diagonal polarizability tensor:
\begin{equation}
E^l_t=\frac{E^l_{b;t}+\alpha^{ll}_{n,eff}G^{ll}_{tn}E^l_{b;n}}{1-\alpha^{ll}_t\{G^{ll}_{tt}+\alpha^{ll}_{n,eff}\left(G^{ll}_{tn}\right)^2 \}}
\label{eq:Etotscalar2}
\end{equation}
\noindent where \(\alpha^{ll}_{n,eff}\), the effective polarizability of the NP, is defined as:
\begin{equation}
\alpha^{ll}_{n,eff}=\frac{\alpha^{ll}_{n}}{1-\alpha^{ll}_{n}G^{ll}_{sc;nn}}
\label{eq:AlphaEff}
\end{equation}
Note that $\alpha^{ll}_{n,eff}$ includes the effects of the self interaction of the NP via the substrate.  The numerator in Eq. (\ref{eq:Etotscalar2}) can be interpreted as a background field which induces the primary dipole moment in the tip, and the denominator \(W^{ll}_{tt}=1-\alpha^{ll}_t\{G^{ll}_{tt}+\alpha^{ll}_{n,eff}\left(G^{ll}_{tn}\right)^2 \}\) accounts for all the multiple scatterings mechanisms that modify the tip local field. We assume that the location of the NP is fixed, implying that \(\alpha^{ll}_{n,eff}\) and \(E_{b;n}^{l}\) are constant numbers in the calculation of the field gradient along \(z\). Under these conditions, we find the following expression for the field gradient along $z$: 

\begin{align} \nonumber
\frac{\partial E^l_t}{\partial z}=\frac{-\gamma E^l_{b;t}+\alpha^{ll}_{n,eff}\frac{\partial G^{ll}_{tn}}{\partial z}E^l_{b;n}}{W^{ll}_{tt}}\qquad\\
+E^l_t\frac{\alpha^{ll}_t\{\frac{\partial G^{ll}_{tt}}{\partial z}+2\alpha^{ll}_{n,eff}\frac{\partial G^{ll}_{tn}}{\partial z}G^{ll}_{tn}\}}{W^{ll}_{tt}}
\label{eq:EtGrad}
\end{align}

\begin{figure}[ht]
	\begin{center}
		\includegraphics[width=0.92\columnwidth]{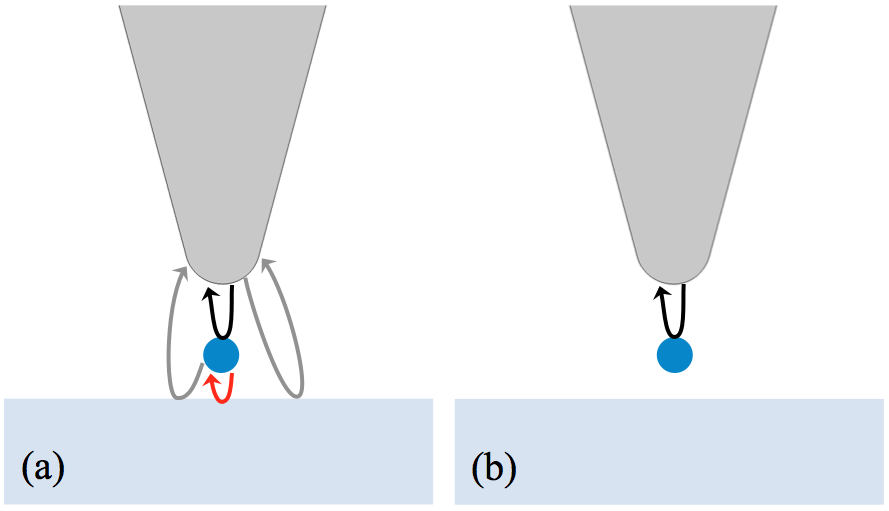}
	\end{center}
	\caption{Schematic of the scattering pathways in the tip-nanoparticle interaction. (a) Pathways including scattering via the substrate, as described by \(\bar{\mathbf{G}}_{sc;tn}\). (b) Direct pathway, as described by \(\bar{\mathbf{G}}_{0}\).} 
	\label{fig:Figure9WtPaths}
\end{figure}

We emphasize that Eqs. (\ref{eq:Etotscalar2}) and (\ref{eq:EtGrad}) include all scattering mechanisms shown in Figure \ref{fig:Figure9WtPaths}(a), accounting for multiple scattering pathways to the tip via the substrate. Our aim is to understand to what extent scattering via the substrate contributes to the overall force experienced by the tip. Therefore, we also consider the dipole-dipole interaction model and define the field in the case of direct scattering between the tip and NP, ignoring the substrate contributions, as depicted in Figure \ref{fig:Figure9WtPaths}(b). Under this condition, Eq.~(\ref{eq:Etotscalar2}) reduces to
\begin{equation}
E^l_t=\frac{E^l_{b;t}+\alpha^{ll}_{n}G^{ll}_{0;tn}E^l_{b;n}}{1-\alpha^{ll}_t\alpha^{ll}_{n}\left(G^{ll}_{0;tn}\right)^2 }
\label{eq:Etotscalar2G0}
\end{equation}

\begin{figure}[ht]
	\begin{center}
		\includegraphics[width=0.92\columnwidth]{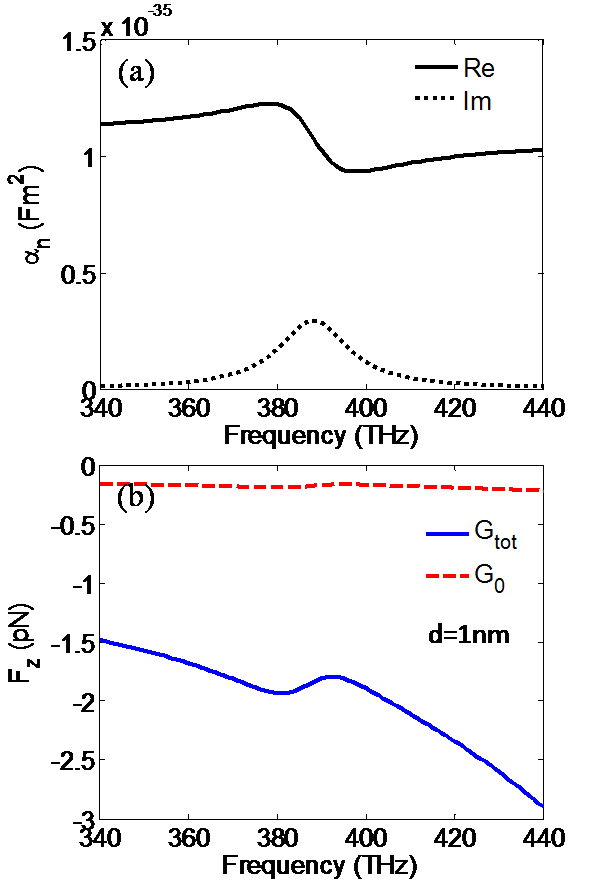}
	\end{center}
	\caption{(a) Spectral dependence of the polarizability of a generic nanoparticle with radius $R_n=4$ nm used in the computation of the photo-induced force. (b) \(F_z\) exerted on the anisotrpic gold tip (AF=2) calculated by including scattering pathways via the substrate (blue solid line) and by ignoring scattering pathways via the substrate (red dashed line). The surface to surface distance between the tip and the NP is 1 nm.}
	\label{fig:FigureForcQD}
\end{figure}

\noindent whose gradient along $z$ can be obtained as: 

\begin{align} \nonumber
\frac{\partial E^l_t}{\partial z}=\frac{-\gamma E^l_{b;t}+\alpha^{ll}_{n}\frac{\partial G^{ll}_{0;tn}}{\partial z}E^l_{b;n}}{W^{ll}_{tt}}\qquad\\
+E^l_t\frac{2\alpha^{ll}_t\alpha^{ll}_{n}\frac{\partial G^{ll}_{0;tn}}{\partial z}G^{ll}_{0;tn}}{W^{ll}_{tt}}
\label{eq:EtG0Grad}
\end{align}

\noindent where the denominator now reads \(W^{ll}_{tt}=1-\alpha^{ll}_t\alpha^{ll}_{n}\left(G^{ll}_{0;tn}\right)^2 \). Using the fields defined in Eqs.~(\ref{eq:Etotscalar2}-\ref{eq:EtGrad}) or (\ref{eq:Etotscalar2G0}-\ref{eq:EtG0Grad}), we can compute calculate the force for the two cases with the aid of Eq. (\ref{eq:force1}). To illustrate the basic features of the photo-induced force in the presence of a nanoparticle, we consider a generic nanoparticle of radius $R_n=4$ nm that is placed on top of a glass substrate, right under the tip, and which exhibits a spectrally dependent polarizability as shown in Figure \ref{fig:FigureForcQD}(a). The background field is evanescent as discussed before. Figure \ref{fig:FigureForcQD}(b) shows the computed photo-induced force in the case of direct scattering (red dashed line) and multiple-scattering with the substrate effect included (blue solid line). It can be seen that the spectral signature of the NP is imprinted on the photo-induced force exerted on the (anisotropic) tip. The spectral dependence of the force follows the dispersive part of the polarizability. The latter is expected when the tip polarizability is dominated by $\alpha^\prime_t$~\cite{Jahng2016spie}, which is the case in this spectral range. This situation is relevant to actual experimental settings, as is reported in~\cite{JahngACR2015}. Figure \ref{fig:FigureForcQD}(b) also makes clear that the influence of the substrate can be substantial. In this particular example, the force is more than ten times stronger when multiple scattering via the substrate is properly accounted for. We find that the substrate effect can be significant, generally increasing the magnitude of the photo-induced force relative to the case of an isolated NP. 

%************************************************************************
%************************************************************************
%************************************************************************
%************************************************************************

\section{\label{sec:Summary}Conclusion}

In this work, we have developed an electrodynamic model to describe the optically induced forces in a tip-sample nanojunction based on dyadic Green's functions for a planar multilayer substrate. Using this description, we have examined the magnitude and behavior of the photo-induced forces in different cases relevant to PiFM experiments. Our analysis has enabled a detailed look at the influence of field and field gradient effects in the dipolar limit, effects that cannot be easily examined independently when using finite element methods.  In addition, our model has allowed us to study the ramifications of scattering pathways via the substrate and the effect of tip anisotropy, and how these mechanisms contribute to the observable force as measured in PiFM, in a physically intuitive and accessible manner.

The qualitative and quantitative observations made with this dyadic Green's function approach go beyond what was learned from simpler models for the photo-induced force. In addition, the model has allowed new and complementary insights to what finite element methods have revealed about photo-induced forces in PiFM type measurements. 

Our analysis has brought forth the following points: 1) The magnitude of the photo-induced force in the tip-sample junction is a very sensitive function of the field gradient. Our analysis shows that multiple scattering effects via the substrate introduce steep gradients that give rise to a significant enhancement of the force. Neglecting multiple-scattering pathways via the substrate thus leads to an underestimation of the force. 2) For a tip with an isotropic polarizability, the photo-induced force over a glass substrate is weak (less than a pN), but can be increased substantially when the substrate is covered with a thin gold film, raising the force into the tens of pN range. 3) The photo-induced force carries a spectral dependence that reflects the polarizability of a nanoparticle under the tip. Our scattering Green's function analysis confirms that PiFM measurements can reveal spectroscopic signatures of particles probed by the tip. 4) Strong photo-induced forces are predicted when the tip anisotropy is included in the analysis. We also find a remarkable sign reversal upon increasing the tip's polarizability, which we attribute to spatial resonance effects in the tip-sample junction. The sign reversal is a unique feature of the photo-induced force, which, unlike many optical signals, sensitively depends on the field gradients in the tip-sample nano-junction. Finally, by including the effects of multiple scattering and tip anisotropy, our model predicts photo-induced forces that approach the same order of magnitude as the forces reported in experimental PiFM studies.

\section*{Acknowlegdments}
This work was supported by the National Science Foundation through grant CHE-1414466.

%************************************************************************
%************************************************************************
%************************************************************************
%************************************************************************

\appendix
\section{Back ground field and force without any scattering mechanisms}

In the following, we assume the substrate/air interface illuminated by plane wave from the substrate side. The resulting background field for $z>0$ can be written as \(\mathbf{E}_b(x,z) = \mathbf{E}_{b0} e^{i(k_xx+k_zz)}\) where \(\mathbf{E}_{b0}\),  \(k_x\) and \(k_z\) are obtained from the Fresnel equations \cite{Novotny2012}. Using Eq.~(\ref{eq:force1}) the photo-induced force exerted by the background field on any isotropically polarizable particle can be expressed as:

\begin{equation}
\mathbf{F}=\frac{|\mathbf{E}_{b}|^2}{2} \Re  \left\{ i\alpha^* (k_x \hat{\mathbf{x}} + k_z \hat{\mathbf{z}})  \right\} 
\label{eq:AppForceb}
\end{equation}
 
Eq.~(\ref{eq:AppForceb}) is valid for both \textit{s} and \textit{p} polarization states of the illumination field, with the exception that \(\mathbf{E}_{b0}\) has to be calculated separately for each polarization state. In case the incident angle is beyond the critical angle, the transmitted field is evanescenct, (\(k_z=i\gamma\)) and the force reads 
 
\begin{equation}
\mathbf{F}=\frac{|\mathbf{E}_{b0}|^2}{2}  ( \alpha^{\prime\prime} k_x\hat{\mathbf{x}} -\alpha^{\prime} \gamma \hat{\mathbf{z}}) e^{-2\gamma z}
\label{eq:AppForceEv}
\end{equation}
 
We see that the $z$-directed force is an attractive force towards the substrate, which is exponentially decaying with \(e^{-2\gamma z}\) away from the interface. In addition, an $x$-directed scattering force is present due to propagation of the surface waves, which decays in the same manner as \(F_z\).  

\section{Anisotropic polarizability}
\label{App:anisotropic}
As discussed in \ref{sec:AnisoPol}, an AFM tip typically exhibits an anisotropic response, which gives rise to a stronger field enhancement in the longitudinal polarization direction of the excitation field ~\cite{Novotny1997,Bouhelier2003,Novotny2006,Novotny2012}. In order to calculate the longitudinal polarizability, we seek to describe the perturbation of the spherical apex toward a virtual prolate. We assume a prolate with a major axis of length $a$ and a minor axis of length $b$. The static polarizability for such a prolate particle in SI units reads~\cite{Link1999,Imura2011}, 

\begin{equation}
\alpha^{ll}=V\varepsilon_0 \varepsilon_h\frac{\varepsilon_t-\varepsilon_h}{L^l(\varepsilon_t-\varepsilon_h)+\varepsilon_h}  
\label{eq:ProlatePol}
\end{equation}
where $V=\frac{4\pi}{3}ab^2 $ is the volume of the prolate, and $L^l$ is the depolarization factor:
\begin{align} \nonumber
L^z=\frac{1-e^2}{e^2}\left(-1+\frac{1}{2e}ln\frac{1+e}{1-e} \right)\quad\\
e^2=1-\frac{1}{AF^2}
\label{eq:ShapeFactors}
\end{align}
Here $AF=a/b$ is the anisotropy factor and 
\begin{equation}
L^x=L^y=\frac{1-L^z}{2}  
\label{eq:LxLy}
\end{equation}

We will also include the effect of dipole radiation damping by writing the anisotropic polarizability of the tip is modeled as:~\cite{Anger2006}
\begin{equation}
\bar{\boldsymbol\alpha}_t=\bar{\boldsymbol\alpha} \left[\bar{\mathbf{I}}-\frac{ik^3}{6\pi\varepsilon_0 \varepsilon_h}\bar{\boldsymbol\alpha}\right]^{-1}
\label{eq:RadLoss}
\end{equation}
 
This model for the polarizability captures the basic features of tip anisotropy, including reasonable estimates of the stronger longitudinal polarizability component and the associated red shift of the its resonance frequency relative to the lateral polarizability components. To enable a direct comparison with the isotropic model, we will assume that the location of the effective induced dipole moment at the tip is the same as in the isotropic model. Although the current model is not intended for definite quantitative predictions of the tip polarizability, we should mention that for the small $AF$ considered in this paper, the predicted numerical values are within practical ranges. The enhancement of the longitudinal polarizability in this model compares well with the enhancement factor reported in the literature~\cite{Bouhelier2003,Novotny2006,Novotny2012}. For instance, values reported in \cite{Novotny2006} based on numerical calculations of the enhancement factor for a 10 nm gold tip at \(\lambda=830 nm\) (far from the resonant frequency) is equivalent to a longitudinal polarizability enhancement, \(|\alpha^{zz}/\alpha^{xx}|\),  of about 11. For $AF=3$ in our treatment, the enhancement of the longitudinal polarizability is around 5 at the red side of the resonance, and is below 10 within the whole frequency range considered here.

%************************************************************************
%************************************************************************
%************************************************************************
%************************************************************************

%\bibliographystyle{unsrt}
\bibliography{biblibrary}

\begin{thebibliography}{39}%
\makeatletter
\providecommand \@ifxundefined [1]{%
 \@ifx{#1\undefined}
}%
\providecommand \@ifnum [1]{%
 \ifnum #1\expandafter \@firstoftwo
 \else \expandafter \@secondoftwo
 \fi
}%
\providecommand \@ifx [1]{%
 \ifx #1\expandafter \@firstoftwo
 \else \expandafter \@secondoftwo
 \fi
}%
\providecommand \natexlab [1]{#1}%
\providecommand \enquote  [1]{``#1''}%
\providecommand \bibnamefont  [1]{#1}%
\providecommand \bibfnamefont [1]{#1}%
\providecommand \citenamefont [1]{#1}%
\providecommand \href@noop [0]{\@secondoftwo}%
\providecommand \href [0]{\begingroup \@sanitize@url \@href}%
\providecommand \@href[1]{\@@startlink{#1}\@@href}%
\providecommand \@@href[1]{\endgroup#1\@@endlink}%
\providecommand \@sanitize@url [0]{\catcode `\\12\catcode `\$12\catcode
  `\&12\catcode `\#12\catcode `\^12\catcode `\_12\catcode `\%12\relax}%
\providecommand \@@startlink[1]{}%
\providecommand \@@endlink[0]{}%
\providecommand \url  [0]{\begingroup\@sanitize@url \@url }%
\providecommand \@url [1]{\endgroup\@href {#1}{\urlprefix }}%
\providecommand \urlprefix  [0]{URL }%
\providecommand \Eprint [0]{\href }%
\providecommand \doibase [0]{http://dx.doi.org/}%
\providecommand \selectlanguage [0]{\@gobble}%
\providecommand \bibinfo  [0]{\@secondoftwo}%
\providecommand \bibfield  [0]{\@secondoftwo}%
\providecommand \translation [1]{[#1]}%
\providecommand \BibitemOpen [0]{}%
\providecommand \bibitemStop [0]{}%
\providecommand \bibitemNoStop [0]{.\EOS\space}%
\providecommand \EOS [0]{\spacefactor3000\relax}%
\providecommand \BibitemShut  [1]{\csname bibitem#1\endcsname}%
\let\auto@bib@innerbib\@empty
%</preamble>
\bibitem [{\citenamefont {Ashkin}(1970)}]{Ashkin1970}%
  \BibitemOpen
  \bibfield  {author} {\bibinfo {author} {\bibfnamefont {A.}~\bibnamefont
  {Ashkin}},\ }\href {\doibase 10.1103/PhysRevLett.24.156} {\bibfield
  {journal} {\bibinfo  {journal} {Phys. Rev. Lett.}\ }\textbf {\bibinfo
  {volume} {24}},\ \bibinfo {pages} {156} (\bibinfo {year} {1970})}\BibitemShut
  {NoStop}%
\bibitem [{\citenamefont {Ashkin}\ and\ \citenamefont
  {Dziedzic}(1987)}]{Ashkin1987}%
  \BibitemOpen
  \bibfield  {author} {\bibinfo {author} {\bibfnamefont {A.}~\bibnamefont
  {Ashkin}}\ and\ \bibinfo {author} {\bibfnamefont {J.}~\bibnamefont
  {Dziedzic}},\ }\href {\doibase 10.1126/science.3547653} {\bibfield  {journal}
  {\bibinfo  {journal} {Science}\ }\textbf {\bibinfo {volume} {235}},\ \bibinfo
  {pages} {1517} (\bibinfo {year} {1987})}\BibitemShut {NoStop}%
\bibitem [{\citenamefont {Dholakia}\ and\ \citenamefont
  {Zem\'anek}(2010)}]{Dholakia2010}%
  \BibitemOpen
  \bibfield  {author} {\bibinfo {author} {\bibfnamefont {K.}~\bibnamefont
  {Dholakia}}\ and\ \bibinfo {author} {\bibfnamefont {P.}~\bibnamefont
  {Zem\'anek}},\ }\href {\doibase 10.1103/RevModPhys.82.1767} {\bibfield
  {journal} {\bibinfo  {journal} {Rev. Mod. Phys.}\ }\textbf {\bibinfo {volume}
  {82}},\ \bibinfo {pages} {1767} (\bibinfo {year} {2010})}\BibitemShut
  {NoStop}%
\bibitem [{\citenamefont {Arias-Gonzalez}\ and\ \citenamefont
  {Nieto-Vesperinas}(2003)}]{AriasGonzalez2003}%
  \BibitemOpen
  \bibfield  {author} {\bibinfo {author} {\bibfnamefont {J.}~\bibnamefont
  {Arias-Gonzalez}}\ and\ \bibinfo {author} {\bibfnamefont {M.}~\bibnamefont
  {Nieto-Vesperinas}},\ }\href@noop {} {\bibfield  {journal} {\bibinfo
  {journal} {J. Opt. Soc. Am. A}\ }\textbf {\bibinfo {volume} {20}},\ \bibinfo
  {pages} {1201} (\bibinfo {year} {2003})}\BibitemShut {NoStop}%
\bibitem [{\citenamefont {Liu}\ \emph {et~al.}(2010)\citenamefont {Liu},
  \citenamefont {Zentgraf}, \citenamefont {Liu}, \citenamefont {Bartal},\ and\
  \citenamefont {Zhang}}]{Liu2010}%
  \BibitemOpen
  \bibfield  {author} {\bibinfo {author} {\bibfnamefont {M.}~\bibnamefont
  {Liu}}, \bibinfo {author} {\bibfnamefont {T.}~\bibnamefont {Zentgraf}},
  \bibinfo {author} {\bibfnamefont {Y.}~\bibnamefont {Liu}}, \bibinfo {author}
  {\bibfnamefont {G.}~\bibnamefont {Bartal}}, \ and\ \bibinfo {author}
  {\bibfnamefont {X.}~\bibnamefont {Zhang}},\ }\href {\doibase
  doi:10.1038/nnano.2010.128} {\bibfield  {journal} {\bibinfo  {journal} {Nat.
  Nanotechnol.}\ }\textbf {\bibinfo {volume} {5}},\ \bibinfo {pages} {570}
  (\bibinfo {year} {2010})}\BibitemShut {NoStop}%
\bibitem [{\citenamefont {Juan}\ \emph {et~al.}(2011)\citenamefont {Juan},
  \citenamefont {Righini},\ and\ \citenamefont {Quidant\"affer}}]{Juan2011}%
  \BibitemOpen
  \bibfield  {author} {\bibinfo {author} {\bibfnamefont {M.~L.}\ \bibnamefont
  {Juan}}, \bibinfo {author} {\bibfnamefont {M.}~\bibnamefont {Righini}}, \
  and\ \bibinfo {author} {\bibfnamefont {R.}~\bibnamefont {Quidant\"affer}},\
  }\href {\doibase doi:10.1038/nphoton.2011.56} {\bibfield  {journal} {\bibinfo
   {journal} {Nat. Photon.}\ }\textbf {\bibinfo {volume} {5}},\ \bibinfo
  {pages} {349} (\bibinfo {year} {2011})}\BibitemShut {NoStop}%
\bibitem [{\citenamefont {Huang}\ \emph {et~al.}(2015)\citenamefont {Huang},
  \citenamefont {Yang}, \citenamefont {Wei}, \citenamefont {Du},\ and\
  \citenamefont {Cui}}]{Huang2015}%
  \BibitemOpen
  \bibfield  {author} {\bibinfo {author} {\bibfnamefont {J.}~\bibnamefont
  {Huang}}, \bibinfo {author} {\bibfnamefont {Z.}~\bibnamefont {Yang}},
  \bibinfo {author} {\bibfnamefont {D.}~\bibnamefont {Wei}}, \bibinfo {author}
  {\bibfnamefont {C.}~\bibnamefont {Du}}, \ and\ \bibinfo {author}
  {\bibfnamefont {H.-L.}\ \bibnamefont {Cui}},\ }\href
  {http://www.mdpi.com/2076-3417/5/4/1745} {\bibfield  {journal} {\bibinfo
  {journal} {Appl. Sci.}\ }\textbf {\bibinfo {volume} {5}},\ \bibinfo {pages}
  {1745} (\bibinfo {year} {2015})}\BibitemShut {NoStop}%
\bibitem [{\citenamefont {Jahng}\ \emph
  {et~al.}(2015{\natexlab{a}})\citenamefont {Jahng}, \citenamefont {Ladani},
  \citenamefont {Khan}, \citenamefont {Li}, \citenamefont {Lee},\ and\
  \citenamefont {Potma}}]{JahngSPP2015}%
  \BibitemOpen
  \bibfield  {author} {\bibinfo {author} {\bibfnamefont {J.}~\bibnamefont
  {Jahng}}, \bibinfo {author} {\bibfnamefont {F.~T.}\ \bibnamefont {Ladani}},
  \bibinfo {author} {\bibfnamefont {R.~M.}\ \bibnamefont {Khan}}, \bibinfo
  {author} {\bibfnamefont {X.}~\bibnamefont {Li}}, \bibinfo {author}
  {\bibfnamefont {E.~S.}\ \bibnamefont {Lee}}, \ and\ \bibinfo {author}
  {\bibfnamefont {E.~O.}\ \bibnamefont {Potma}},\ }\href {\doibase
  10.1364/OL.40.005058} {\bibfield  {journal} {\bibinfo  {journal} {Opt.
  Lett.}\ }\textbf {\bibinfo {volume} {40}},\ \bibinfo {pages} {5058} (\bibinfo
  {year} {2015}{\natexlab{a}})}\BibitemShut {NoStop}%
\bibitem [{\citenamefont {Kohoutek}\ \emph {et~al.}(2011)\citenamefont
  {Kohoutek}, \citenamefont {Dey}, \citenamefont {Bonakdar}, \citenamefont
  {Gelfand}, \citenamefont {Sklar}, \citenamefont {Memis},\ and\ \citenamefont
  {Mohseni}}]{Kohoutek2011}%
  \BibitemOpen
  \bibfield  {author} {\bibinfo {author} {\bibfnamefont {J.}~\bibnamefont
  {Kohoutek}}, \bibinfo {author} {\bibfnamefont {D.}~\bibnamefont {Dey}},
  \bibinfo {author} {\bibfnamefont {A.}~\bibnamefont {Bonakdar}}, \bibinfo
  {author} {\bibfnamefont {R.}~\bibnamefont {Gelfand}}, \bibinfo {author}
  {\bibfnamefont {A.}~\bibnamefont {Sklar}}, \bibinfo {author} {\bibfnamefont
  {O.~G.}\ \bibnamefont {Memis}}, \ and\ \bibinfo {author} {\bibfnamefont
  {H.}~\bibnamefont {Mohseni}},\ }\href {\doibase 10.1021/nl201780y} {\bibfield
   {journal} {\bibinfo  {journal} {Nano Lett.}\ }\textbf {\bibinfo {volume}
  {11}},\ \bibinfo {pages} {3378} (\bibinfo {year} {2011})}\BibitemShut
  {NoStop}%
\bibitem [{\citenamefont {Tumkur}\ \emph {et~al.}(2016)\citenamefont {Tumkur},
  \citenamefont {Yang}, \citenamefont {Cerjan}, \citenamefont {Halas},
  \citenamefont {Nordlander},\ and\ \citenamefont {Thomann}}]{Tumkur2016}%
  \BibitemOpen
  \bibfield  {author} {\bibinfo {author} {\bibfnamefont {T.~U.}\ \bibnamefont
  {Tumkur}}, \bibinfo {author} {\bibfnamefont {X.}~\bibnamefont {Yang}},
  \bibinfo {author} {\bibfnamefont {B.}~\bibnamefont {Cerjan}}, \bibinfo
  {author} {\bibfnamefont {N.~J.}\ \bibnamefont {Halas}}, \bibinfo {author}
  {\bibfnamefont {P.}~\bibnamefont {Nordlander}}, \ and\ \bibinfo {author}
  {\bibfnamefont {I.}~\bibnamefont {Thomann}},\ }\href {\doibase
  10.1021/acs.nanolett.6b04245} {\bibfield  {journal} {\bibinfo  {journal}
  {Nano Lett.}\ }\textbf {\bibinfo {volume} {16}},\ \bibinfo {pages} {7942}
  (\bibinfo {year} {2016})}\BibitemShut {NoStop}%
\bibitem [{\citenamefont {Rajapaksa}\ \emph {et~al.}(2010)\citenamefont
  {Rajapaksa}, \citenamefont {Uenal},\ and\ \citenamefont
  {Wickramasinghe}}]{Rajapaksa2010}%
  \BibitemOpen
  \bibfield  {author} {\bibinfo {author} {\bibfnamefont {I.}~\bibnamefont
  {Rajapaksa}}, \bibinfo {author} {\bibfnamefont {K.}~\bibnamefont {Uenal}}, \
  and\ \bibinfo {author} {\bibfnamefont {H.~K.}\ \bibnamefont
  {Wickramasinghe}},\ }\href@noop {} {\bibfield  {journal} {\bibinfo  {journal}
  {Appl. Phys. Lett.}\ }\textbf {\bibinfo {volume} {97}},\ \bibinfo {eid}
  {073121} (\bibinfo {year} {2010})}\BibitemShut {NoStop}%
\bibitem [{\citenamefont {Rajapaksa}\ and\ \citenamefont
  {Kumar~Wickramasinghe}(2011)}]{Rajapaksa2011}%
  \BibitemOpen
  \bibfield  {author} {\bibinfo {author} {\bibfnamefont {I.}~\bibnamefont
  {Rajapaksa}}\ and\ \bibinfo {author} {\bibfnamefont {H.}~\bibnamefont
  {Kumar~Wickramasinghe}},\ }\href@noop {} {\bibfield  {journal} {\bibinfo
  {journal} {Appl. Phys. Lett.}\ }\textbf {\bibinfo {volume} {99}},\ \bibinfo
  {eid} {161103} (\bibinfo {year} {2011})}\BibitemShut {NoStop}%
\bibitem [{\citenamefont {Jahng}\ \emph
  {et~al.}(2015{\natexlab{b}})\citenamefont {Jahng}, \citenamefont {Brocious},
  \citenamefont {Fishman}, \citenamefont {Yampolsky}, \citenamefont {Nowak},
  \citenamefont {Huang}, \citenamefont {Apkarian}, \citenamefont
  {Wickramasinghe},\ and\ \citenamefont {Potma}}]{JahngUltra2015}%
  \BibitemOpen
  \bibfield  {author} {\bibinfo {author} {\bibfnamefont {J.}~\bibnamefont
  {Jahng}}, \bibinfo {author} {\bibfnamefont {J.}~\bibnamefont {Brocious}},
  \bibinfo {author} {\bibfnamefont {D.~A.}\ \bibnamefont {Fishman}}, \bibinfo
  {author} {\bibfnamefont {S.}~\bibnamefont {Yampolsky}}, \bibinfo {author}
  {\bibfnamefont {D.}~\bibnamefont {Nowak}}, \bibinfo {author} {\bibfnamefont
  {F.}~\bibnamefont {Huang}}, \bibinfo {author} {\bibfnamefont {V.~A.}\
  \bibnamefont {Apkarian}}, \bibinfo {author} {\bibfnamefont {H.~K.}\
  \bibnamefont {Wickramasinghe}}, \ and\ \bibinfo {author} {\bibfnamefont
  {E.~O.}\ \bibnamefont {Potma}},\ }\href {http://dx.doi.org/10.1063/1.4913853}
  {\bibfield  {journal} {\bibinfo  {journal} {Appl. Phys. Lett.}\ }\textbf
  {\bibinfo {volume} {106}},\ \bibinfo {eid} {083113} (\bibinfo {year}
  {2015}{\natexlab{b}})}\BibitemShut {NoStop}%
\bibitem [{\citenamefont {Jahng}\ \emph {et~al.}()\citenamefont {Jahng},
  \citenamefont {Fishman}, \citenamefont {Park}, \citenamefont {Nowak},
  \citenamefont {Morrison}, \citenamefont {Wickramasinghe},\ and\ \citenamefont
  {Potma}}]{JahngACR2015}%
  \BibitemOpen
  \bibfield  {author} {\bibinfo {author} {\bibfnamefont {J.}~\bibnamefont
  {Jahng}}, \bibinfo {author} {\bibfnamefont {D.~A.}\ \bibnamefont {Fishman}},
  \bibinfo {author} {\bibfnamefont {S.}~\bibnamefont {Park}}, \bibinfo {author}
  {\bibfnamefont {D.~B.}\ \bibnamefont {Nowak}}, \bibinfo {author}
  {\bibfnamefont {W.~A.}\ \bibnamefont {Morrison}}, \bibinfo {author}
  {\bibfnamefont {H.~K.}\ \bibnamefont {Wickramasinghe}}, \ and\ \bibinfo
  {author} {\bibfnamefont {E.~O.}\ \bibnamefont {Potma}},\ }\href@noop {} {\
  }\BibitemShut {NoStop}%
\bibitem [{\citenamefont {Nowak}\ \emph {et~al.}(2016)\citenamefont {Nowak},
  \citenamefont {Morrison}, \citenamefont {Wickramasinghe}, \citenamefont
  {Jahng}, \citenamefont {Potma}, \citenamefont {Wan}, \citenamefont {Ruiz},
  \citenamefont {Albrecht}, \citenamefont {Schmidt}, \citenamefont {Frommer},
  \citenamefont {Sanders},\ and\ \citenamefont {Park}}]{Nowake2016}%
  \BibitemOpen
  \bibfield  {author} {\bibinfo {author} {\bibfnamefont {D.}~\bibnamefont
  {Nowak}}, \bibinfo {author} {\bibfnamefont {W.}~\bibnamefont {Morrison}},
  \bibinfo {author} {\bibfnamefont {H.~K.}\ \bibnamefont {Wickramasinghe}},
  \bibinfo {author} {\bibfnamefont {J.}~\bibnamefont {Jahng}}, \bibinfo
  {author} {\bibfnamefont {E.}~\bibnamefont {Potma}}, \bibinfo {author}
  {\bibfnamefont {L.}~\bibnamefont {Wan}}, \bibinfo {author} {\bibfnamefont
  {R.}~\bibnamefont {Ruiz}}, \bibinfo {author} {\bibfnamefont {T.~R.}\
  \bibnamefont {Albrecht}}, \bibinfo {author} {\bibfnamefont {K.}~\bibnamefont
  {Schmidt}}, \bibinfo {author} {\bibfnamefont {J.}~\bibnamefont {Frommer}},
  \bibinfo {author} {\bibfnamefont {D.~P.}\ \bibnamefont {Sanders}}, \ and\
  \bibinfo {author} {\bibfnamefont {S.}~\bibnamefont {Park}},\ }\href
  {http://advances.sciencemag.org/content/2/3/e1501571} {\bibfield  {journal}
  {\bibinfo  {journal} {Sci. Adv.}\ }\textbf {\bibinfo {volume} {2}} (\bibinfo
  {year} {2016})}\BibitemShut {NoStop}%
\bibitem [{\citenamefont {Jahng}\ \emph {et~al.}(2014)\citenamefont {Jahng},
  \citenamefont {Brocious}, \citenamefont {Fishman}, \citenamefont {Huang},
  \citenamefont {Li}, \citenamefont {Tamma}, \citenamefont {Wickramasinghe},\
  and\ \citenamefont {Potma}}]{Jahng2014}%
  \BibitemOpen
  \bibfield  {author} {\bibinfo {author} {\bibfnamefont {J.}~\bibnamefont
  {Jahng}}, \bibinfo {author} {\bibfnamefont {J.}~\bibnamefont {Brocious}},
  \bibinfo {author} {\bibfnamefont {D.~A.}\ \bibnamefont {Fishman}}, \bibinfo
  {author} {\bibfnamefont {F.}~\bibnamefont {Huang}}, \bibinfo {author}
  {\bibfnamefont {X.}~\bibnamefont {Li}}, \bibinfo {author} {\bibfnamefont
  {V.~A.}\ \bibnamefont {Tamma}}, \bibinfo {author} {\bibfnamefont {H.~K.}\
  \bibnamefont {Wickramasinghe}}, \ and\ \bibinfo {author} {\bibfnamefont
  {E.~O.}\ \bibnamefont {Potma}},\ }\href {\doibase 10.1103/PhysRevB.90.155417}
  {\bibfield  {journal} {\bibinfo  {journal} {Phys. Rev. B}\ }\textbf {\bibinfo
  {volume} {90}},\ \bibinfo {pages} {155417} (\bibinfo {year}
  {2014})}\BibitemShut {NoStop}%
\bibitem [{\citenamefont {Jahng}\ \emph
  {et~al.}(2016{\natexlab{a}})\citenamefont {Jahng}, \citenamefont {Kim},
  \citenamefont {Lee},\ and\ \citenamefont {Potma}}]{Jahng2016PRB}%
  \BibitemOpen
  \bibfield  {author} {\bibinfo {author} {\bibfnamefont {J.}~\bibnamefont
  {Jahng}}, \bibinfo {author} {\bibfnamefont {B.}~\bibnamefont {Kim}}, \bibinfo
  {author} {\bibfnamefont {E.~S.}\ \bibnamefont {Lee}}, \ and\ \bibinfo
  {author} {\bibfnamefont {E.~O.}\ \bibnamefont {Potma}},\ }\href {\doibase
  10.1103/PhysRevB.94.195407} {\bibfield  {journal} {\bibinfo  {journal} {Phys.
  Rev. B}\ }\textbf {\bibinfo {volume} {94}},\ \bibinfo {pages} {195407}
  (\bibinfo {year} {2016}{\natexlab{a}})}\BibitemShut {NoStop}%
\bibitem [{\citenamefont {Novotny}\ and\ \citenamefont
  {Hecht}(2012)}]{Novotny2012}%
  \BibitemOpen
  \bibfield  {author} {\bibinfo {author} {\bibfnamefont {N.}~\bibnamefont
  {Novotny}}\ and\ \bibinfo {author} {\bibfnamefont {B.}~\bibnamefont
  {Hecht}},\ }\href@noop {} {\emph {\bibinfo {title} {{Principles of
  Nano-Optics}}}}\ (\bibinfo  {publisher} {Cambridge University Press},\
  \bibinfo {address} {Cambridge, UK},\ \bibinfo {year} {2012})\BibitemShut
  {NoStop}%
\bibitem [{\citenamefont {Yang}\ and\ \citenamefont
  {Raschke}(2016)}]{Yang2016Raschke}%
  \BibitemOpen
  \bibfield  {author} {\bibinfo {author} {\bibfnamefont {H.~U.}\ \bibnamefont
  {Yang}}\ and\ \bibinfo {author} {\bibfnamefont {M.~B.}\ \bibnamefont
  {Raschke}},\ }\href {http://stacks.iop.org/1367-2630/18/i=5/a=053042}
  {\bibfield  {journal} {\bibinfo  {journal} {New J. Phys.}\ }\textbf {\bibinfo
  {volume} {18}},\ \bibinfo {pages} {053042} (\bibinfo {year}
  {2016})}\BibitemShut {NoStop}%
\bibitem [{\citenamefont {Chaumet}\ and\ \citenamefont
  {Nieto-Vesperinas}(2000)}]{Chaumet2000}%
  \BibitemOpen
  \bibfield  {author} {\bibinfo {author} {\bibfnamefont {P.~C.}\ \bibnamefont
  {Chaumet}}\ and\ \bibinfo {author} {\bibfnamefont {M.}~\bibnamefont
  {Nieto-Vesperinas}},\ }\href {\doibase 10.1364/OL.25.001065} {\bibfield
  {journal} {\bibinfo  {journal} {Opt. Lett.}\ }\textbf {\bibinfo {volume}
  {25}},\ \bibinfo {pages} {1065} (\bibinfo {year} {2000})}\BibitemShut
  {NoStop}%
\bibitem [{\citenamefont {Nieto-Vesperinas}\ \emph {et~al.}(2004)\citenamefont
  {Nieto-Vesperinas}, \citenamefont {Chaumet},\ and\ \citenamefont
  {Rahmani}}]{NietoVesperinas2004}%
  \BibitemOpen
  \bibfield  {author} {\bibinfo {author} {\bibfnamefont {M.}~\bibnamefont
  {Nieto-Vesperinas}}, \bibinfo {author} {\bibfnamefont {P.~C.}\ \bibnamefont
  {Chaumet}}, \ and\ \bibinfo {author} {\bibfnamefont {A.}~\bibnamefont
  {Rahmani}},\ }\href {\doibase 10.1098/rsta.2003.1343} {\bibfield  {journal}
  {\bibinfo  {journal} {Phil. Trans. R. Soc. Lond. A}\ }\textbf {\bibinfo
  {volume} {362}},\ \bibinfo {pages} {719} (\bibinfo {year}
  {2004})}\BibitemShut {NoStop}%
\bibitem [{\citenamefont {Jahng}\ \emph
  {et~al.}(2016{\natexlab{b}})\citenamefont {Jahng}, \citenamefont
  {Tork~Ladani}, \citenamefont {Khan},\ and\ \citenamefont
  {Potma}}]{Jahng2016spie}%
  \BibitemOpen
  \bibfield  {author} {\bibinfo {author} {\bibfnamefont {J.}~\bibnamefont
  {Jahng}}, \bibinfo {author} {\bibfnamefont {F.}~\bibnamefont {Tork~Ladani}},
  \bibinfo {author} {\bibfnamefont {R.~M.}\ \bibnamefont {Khan}}, \ and\
  \bibinfo {author} {\bibfnamefont {E.~O.}\ \bibnamefont {Potma}},\ }\href
  {\doibase 10.1117/12.2208199} {\bibfield  {journal} {\bibinfo  {journal}
  {Proc. SPIE}\ }\textbf {\bibinfo {volume} {9764}},\ \bibinfo {pages}
  {97641J1} (\bibinfo {year} {2016}{\natexlab{b}})}\BibitemShut {NoStop}%
\bibitem [{\citenamefont {Zhang}\ \emph {et~al.}(2015)\citenamefont {Zhang},
  \citenamefont {Chen},\ and\ \citenamefont {Li}}]{Zhang2015}%
  \BibitemOpen
  \bibfield  {author} {\bibinfo {author} {\bibfnamefont {C.}~\bibnamefont
  {Zhang}}, \bibinfo {author} {\bibfnamefont {B.-Q.}\ \bibnamefont {Chen}}, \
  and\ \bibinfo {author} {\bibfnamefont {Z.-Y.}\ \bibnamefont {Li}},\ }\href
  {\doibase 10.1021/acs.jpcc.5b02653} {\bibfield  {journal} {\bibinfo
  {journal} {J. Phys. Chem. C}\ }\textbf {\bibinfo {volume} {119}},\ \bibinfo
  {pages} {11858} (\bibinfo {year} {2015})}\BibitemShut {NoStop}%
\bibitem [{\citenamefont {Salary}\ and\ \citenamefont
  {Mosallaei}(2016)}]{Salary2016}%
  \BibitemOpen
  \bibfield  {author} {\bibinfo {author} {\bibfnamefont {M.~M.}\ \bibnamefont
  {Salary}}\ and\ \bibinfo {author} {\bibfnamefont {H.}~\bibnamefont
  {Mosallaei}},\ }\href {\doibase 10.1103/PhysRevB.94.035410} {\bibfield
  {journal} {\bibinfo  {journal} {Phys. Rev. B}\ }\textbf {\bibinfo {volume}
  {94}},\ \bibinfo {pages} {035410} (\bibinfo {year} {2016})}\BibitemShut
  {NoStop}%
\bibitem [{\citenamefont {Kong}(1990)}]{Kong1990}%
  \BibitemOpen
  \bibfield  {author} {\bibinfo {author} {\bibfnamefont {J.~A.}\ \bibnamefont
  {Kong}},\ }\href@noop {} {\emph {\bibinfo {title} {{ Electromagnetic wave
  theory}}}}\ (\bibinfo  {publisher} {Wiley},\ \bibinfo {address} {New York},\
  \bibinfo {year} {1990})\BibitemShut {NoStop}%
\bibitem [{\citenamefont {Novotny}\ \emph {et~al.}(1998)\citenamefont
  {Novotny}, \citenamefont {Sánchez},\ and\ \citenamefont
  {Xie}}]{Novotny1998}%
  \BibitemOpen
  \bibfield  {author} {\bibinfo {author} {\bibfnamefont {L.}~\bibnamefont
  {Novotny}}, \bibinfo {author} {\bibfnamefont {E.~J.}\ \bibnamefont
  {Sánchez}}, \ and\ \bibinfo {author} {\bibfnamefont {X.~S.}\ \bibnamefont
  {Xie}},\ }\href {\doibase http://dx.doi.org/10.1016/S0304-3991(97)00077-6}
  {\bibfield  {journal} {\bibinfo  {journal} {Ultramicroscopy}\ }\textbf
  {\bibinfo {volume} {71}},\ \bibinfo {pages} {21} (\bibinfo {year}
  {1998})}\BibitemShut {NoStop}%
\bibitem [{\citenamefont {Ichimura}\ \emph {et~al.}(2007)\citenamefont
  {Ichimura}, \citenamefont {Watanabe}, \citenamefont {Morita}, \citenamefont
  {Verma}, \citenamefont {Kawata},\ and\ \citenamefont
  {Inouye}}]{Ichimura2007}%
  \BibitemOpen
  \bibfield  {author} {\bibinfo {author} {\bibfnamefont {T.}~\bibnamefont
  {Ichimura}}, \bibinfo {author} {\bibfnamefont {H.}~\bibnamefont {Watanabe}},
  \bibinfo {author} {\bibfnamefont {Y.}~\bibnamefont {Morita}}, \bibinfo
  {author} {\bibfnamefont {P.}~\bibnamefont {Verma}}, \bibinfo {author}
  {\bibfnamefont {S.}~\bibnamefont {Kawata}}, \ and\ \bibinfo {author}
  {\bibfnamefont {Y.}~\bibnamefont {Inouye}},\ }\href {\doibase
  10.1021/jp070420b} {\bibfield  {journal} {\bibinfo  {journal} {J. Phys. Chem.
  C}\ }\textbf {\bibinfo {volume} {111}},\ \bibinfo {pages} {9460} (\bibinfo
  {year} {2007})}\BibitemShut {NoStop}%
\bibitem [{\citenamefont {Olmon}\ \emph {et~al.}(2012)\citenamefont {Olmon},
  \citenamefont {Slovick}, \citenamefont {Johnson}, \citenamefont {Shelton},
  \citenamefont {Oh}, \citenamefont {Boreman},\ and\ \citenamefont
  {Raschke}}]{Olmon2012}%
  \BibitemOpen
  \bibfield  {author} {\bibinfo {author} {\bibfnamefont {R.~L.}\ \bibnamefont
  {Olmon}}, \bibinfo {author} {\bibfnamefont {B.}~\bibnamefont {Slovick}},
  \bibinfo {author} {\bibfnamefont {T.~W.}\ \bibnamefont {Johnson}}, \bibinfo
  {author} {\bibfnamefont {D.}~\bibnamefont {Shelton}}, \bibinfo {author}
  {\bibfnamefont {S.-H.}\ \bibnamefont {Oh}}, \bibinfo {author} {\bibfnamefont
  {G.~D.}\ \bibnamefont {Boreman}}, \ and\ \bibinfo {author} {\bibfnamefont
  {M.~B.}\ \bibnamefont {Raschke}},\ }\href {\doibase
  10.1103/PhysRevB.86.235147} {\bibfield  {journal} {\bibinfo  {journal} {Phys.
  Rev. B}\ }\textbf {\bibinfo {volume} {86}},\ \bibinfo {pages} {235147}
  (\bibinfo {year} {2012})}\BibitemShut {NoStop}%
\bibitem [{\citenamefont {Bohren}\ and\ \citenamefont
  {Huffman}(1998)}]{Bohren1998}%
  \BibitemOpen
  \bibfield  {author} {\bibinfo {author} {\bibfnamefont {C.~F.}\ \bibnamefont
  {Bohren}}\ and\ \bibinfo {author} {\bibfnamefont {D.~R.}\ \bibnamefont
  {Huffman}},\ }\href@noop {} {\emph {\bibinfo {title} {{Absorption and
  Scattering of Light by Small Particles}}}}\ (\bibinfo  {publisher} {Wiley-VCH
  Verlag GmbH},\ \bibinfo {address} {Berlin},\ \bibinfo {year}
  {1998})\BibitemShut {NoStop}%
\bibitem [{\citenamefont {Campione}\ \emph {et~al.}(2011)\citenamefont
  {Campione}, \citenamefont {Albani},\ and\ \citenamefont
  {Capolino}}]{Campione2011}%
  \BibitemOpen
  \bibfield  {author} {\bibinfo {author} {\bibfnamefont {S.}~\bibnamefont
  {Campione}}, \bibinfo {author} {\bibfnamefont {M.}~\bibnamefont {Albani}}, \
  and\ \bibinfo {author} {\bibfnamefont {F.}~\bibnamefont {Capolino}},\ }\href
  {\doibase 10.1364/OME.1.001077} {\bibfield  {journal} {\bibinfo  {journal}
  {Opt. Mater. Express}\ }\textbf {\bibinfo {volume} {1}},\ \bibinfo {pages}
  {1077} (\bibinfo {year} {2011})}\BibitemShut {NoStop}%
\bibitem [{\citenamefont {McPeak}\ \emph {et~al.}(2015)\citenamefont {McPeak},
  \citenamefont {Jayanti}, \citenamefont {Kress}, \citenamefont {Meyer},
  \citenamefont {Iotti}, \citenamefont {Rossinelli},\ and\ \citenamefont
  {Norris}}]{McPeak2015}%
  \BibitemOpen
  \bibfield  {author} {\bibinfo {author} {\bibfnamefont {K.~M.}\ \bibnamefont
  {McPeak}}, \bibinfo {author} {\bibfnamefont {S.~V.}\ \bibnamefont {Jayanti}},
  \bibinfo {author} {\bibfnamefont {S.~J.~P.}\ \bibnamefont {Kress}}, \bibinfo
  {author} {\bibfnamefont {S.}~\bibnamefont {Meyer}}, \bibinfo {author}
  {\bibfnamefont {S.}~\bibnamefont {Iotti}}, \bibinfo {author} {\bibfnamefont
  {A.}~\bibnamefont {Rossinelli}}, \ and\ \bibinfo {author} {\bibfnamefont
  {D.~J.}\ \bibnamefont {Norris}},\ }\href {\doibase 10.1021/ph5004237}
  {\bibfield  {journal} {\bibinfo  {journal} {ACS Photon.}\ }\textbf {\bibinfo
  {volume} {2}},\ \bibinfo {pages} {326} (\bibinfo {year} {2015})}\BibitemShut
  {NoStop}%
\bibitem [{\citenamefont {Tork~Ladani}\ \emph {et~al.}(2014)\citenamefont
  {Tork~Ladani}, \citenamefont {Campione}, \citenamefont {Guclu},\ and\
  \citenamefont {Capolino}}]{FTL2014}%
  \BibitemOpen
  \bibfield  {author} {\bibinfo {author} {\bibfnamefont {F.}~\bibnamefont
  {Tork~Ladani}}, \bibinfo {author} {\bibfnamefont {S.}~\bibnamefont
  {Campione}}, \bibinfo {author} {\bibfnamefont {C.}~\bibnamefont {Guclu}}, \
  and\ \bibinfo {author} {\bibfnamefont {F.}~\bibnamefont {Capolino}},\ }\href
  {\doibase 10.1103/PhysRevB.90.125127} {\bibfield  {journal} {\bibinfo
  {journal} {Phys. Rev. B}\ }\textbf {\bibinfo {volume} {90}},\ \bibinfo
  {pages} {125127} (\bibinfo {year} {2014})}\BibitemShut {NoStop}%
\bibitem [{\citenamefont {Wickramasinghe}\ and\ \citenamefont
  {Rajapaksa}(2011)}]{Wickramasinghe2011}%
  \BibitemOpen
  \bibfield  {author} {\bibinfo {author} {\bibfnamefont {H.}~\bibnamefont
  {Wickramasinghe}}\ and\ \bibinfo {author} {\bibfnamefont {I.}~\bibnamefont
  {Rajapaksa}},\ }\href {\doibase http://dx.doi.org/doi: 10.1557/opl.2011.283}
  {\bibfield  {journal} {\bibinfo  {journal} {MRS Proceedings}\ }\textbf
  {\bibinfo {volume} {1318}} (\bibinfo {year} {2011}),\ http://dx.doi.org/doi:
  10.1557/opl.2011.283}\BibitemShut {NoStop}%
\bibitem [{\citenamefont {Novotny}\ \emph {et~al.}(1997)\citenamefont
  {Novotny}, \citenamefont {Bian},\ and\ \citenamefont {Xie}}]{Novotny1997}%
  \BibitemOpen
  \bibfield  {author} {\bibinfo {author} {\bibfnamefont {L.}~\bibnamefont
  {Novotny}}, \bibinfo {author} {\bibfnamefont {R.~X.}\ \bibnamefont {Bian}}, \
  and\ \bibinfo {author} {\bibfnamefont {X.~S.}\ \bibnamefont {Xie}},\ }\href
  {\doibase 10.1103/PhysRevLett.79.645} {\bibfield  {journal} {\bibinfo
  {journal} {Phys. Rev. Lett.}\ }\textbf {\bibinfo {volume} {79}},\ \bibinfo
  {pages} {645} (\bibinfo {year} {1997})}\BibitemShut {NoStop}%
\bibitem [{\citenamefont {Bouhelier}\ \emph {et~al.}(2003)\citenamefont
  {Bouhelier}, \citenamefont {Beversluis}, \citenamefont {Hartschuh},\ and\
  \citenamefont {Novotny}}]{Bouhelier2003}%
  \BibitemOpen
  \bibfield  {author} {\bibinfo {author} {\bibfnamefont {A.}~\bibnamefont
  {Bouhelier}}, \bibinfo {author} {\bibfnamefont {M.}~\bibnamefont
  {Beversluis}}, \bibinfo {author} {\bibfnamefont {A.}~\bibnamefont
  {Hartschuh}}, \ and\ \bibinfo {author} {\bibfnamefont {L.}~\bibnamefont
  {Novotny}},\ }\href {\doibase 10.1103/PhysRevLett.90.013903} {\bibfield
  {journal} {\bibinfo  {journal} {Phys. Rev. Lett.}\ }\textbf {\bibinfo
  {volume} {90}},\ \bibinfo {pages} {013903} (\bibinfo {year}
  {2003})}\BibitemShut {NoStop}%
\bibitem [{\citenamefont {Novotny}\ and\ \citenamefont
  {Stranick}(2006)}]{Novotny2006}%
  \BibitemOpen
  \bibfield  {author} {\bibinfo {author} {\bibfnamefont {L.}~\bibnamefont
  {Novotny}}\ and\ \bibinfo {author} {\bibfnamefont {S.~J.}\ \bibnamefont
  {Stranick}},\ }\href {\doibase 10.1146/annurev.physchem.56.092503.141236}
  {\bibfield  {journal} {\bibinfo  {journal} {Annu. Rev. Phys. Chem.}\ }\textbf
  {\bibinfo {volume} {57}},\ \bibinfo {pages} {303} (\bibinfo {year}
  {2006})}\BibitemShut {NoStop}%
\bibitem [{\citenamefont {Link}\ \emph {et~al.}(1999)\citenamefont {Link},
  \citenamefont {Mohamed},\ and\ \citenamefont {El-Sayed}}]{Link1999}%
  \BibitemOpen
  \bibfield  {author} {\bibinfo {author} {\bibfnamefont {S.}~\bibnamefont
  {Link}}, \bibinfo {author} {\bibfnamefont {M.~B.}\ \bibnamefont {Mohamed}}, \
  and\ \bibinfo {author} {\bibfnamefont {M.~A.}\ \bibnamefont {El-Sayed}},\
  }\href {\doibase 10.1021/jp990183f} {\bibfield  {journal} {\bibinfo
  {journal} {The Journal of Physical Chemistry B}\ }\textbf {\bibinfo {volume}
  {103}},\ \bibinfo {pages} {3073} (\bibinfo {year} {1999})}\BibitemShut
  {NoStop}%
\bibitem [{\citenamefont {Imura}\ and\ \citenamefont
  {Okamoto}(2011)}]{Imura2011}%
  \BibitemOpen
  \bibfield  {author} {\bibinfo {author} {\bibfnamefont {K.}~\bibnamefont
  {Imura}}\ and\ \bibinfo {author} {\bibfnamefont {H.}~\bibnamefont
  {Okamoto}},\ }in\ \href@noop {} {\emph {\bibinfo {booktitle} {Progress in
  Nanophotonics 1}}},\ \bibinfo {editor} {edited by\ \bibinfo {editor}
  {\bibfnamefont {M.}~\bibnamefont {Ohtsu}}}\ (\bibinfo  {publisher}
  {Springer},\ \bibinfo {address} {Berlin},\ \bibinfo {year} {2011})\ pp.\
  \bibinfo {pages} {127--160}\BibitemShut {NoStop}%
\bibitem [{\citenamefont {Anger}\ \emph {et~al.}(2006)\citenamefont {Anger},
  \citenamefont {Bharadwaj},\ and\ \citenamefont {Novotny}}]{Anger2006}%
  \BibitemOpen
  \bibfield  {author} {\bibinfo {author} {\bibfnamefont {P.}~\bibnamefont
  {Anger}}, \bibinfo {author} {\bibfnamefont {P.}~\bibnamefont {Bharadwaj}}, \
  and\ \bibinfo {author} {\bibfnamefont {L.}~\bibnamefont {Novotny}},\ }\href
  {\doibase 10.1103/PhysRevLett.96.113002} {\bibfield  {journal} {\bibinfo
  {journal} {Phys. Rev. Lett.}\ }\textbf {\bibinfo {volume} {96}},\ \bibinfo
  {pages} {113002} (\bibinfo {year} {2006})}\BibitemShut {NoStop}%
\end{thebibliography}%

\end{document}